\documentclass[12pt]{article} %[reprint,aps]{revtex4-1}
\usepackage{epsfig, amsmath, amssymb}
\usepackage{graphicx,psfrag} 
\newcommand{\be}{\begin{equation}}
\newcommand{\ee}{\end{equation}}
\newcommand{\ben}{\begin{equation}}
\newcommand{\een}{\end{equation}}
\newcommand{\bea}{\begin{eqnarray}}
\newcommand{\eea}{\end{eqnarray}}
\newcommand{\bA}{\begin{array}}
\newcommand{\eA}{\end{array}}
\newcommand{\bc}{\begin{center}}
\newcommand{\ec}{\end{center}}
\newcommand{\al}{\alpha}

\newcommand{\ra}{\rightarrow}
\newcommand{\del}{\partial}

\newcommand{\ie}{{\it i.e.}}
\newcommand{\eg}{{\it e.g.}}

\def\BC{{\mathbb C}}

\def\BR{{\mathbb R}}
\def\BZ{{\mathbb Z}}

\def\BI{{\mathbb I}}

\begin{document}

%\ifeprint
%\fi

\begin{titlepage}
\vspace{30mm}

\bc

%%\hfill  {TIFR/TH/09-12} \\
\hfill % {\tt arXiv:0909.4731 [hep-th]} 
\\         [30mm]
%%X\vfill

%{\Huge On $dS_4$ extremal surfaces\\ 
%[-2mm] 
%and entanglement entropy\\ [3mm] in some ghost CFTs} 
{\Huge On $dS_4$ extremal surfaces and\\ 
[2mm] 
entanglement entropy in some ghost CFTs} 

\vspace{16mm}

{\large K.~Narayan} \\
\vspace{3mm}
{\small \it Chennai Mathematical Institute, \\}
{\small \it SIPCOT IT Park, Siruseri 603103, India.\\}
%%{\small Email: \ narayan@cmi.ac.in}\\

\ec
%\medskip
\vspace{30mm}

\begin{abstract}
In arXiv:1501.03019 [hep-th], the areas of certain complex extremal
surfaces in de Sitter space were found to have resemblance with
entanglement entropy in appropriate dual Euclidean non-unitary CFTs,
with the area being real and negative in $dS_4$. In this paper, we
study some toy models of 2-dim ghost conformal field theories with
negative central charge with a view to exploring this further from the
CFT point of view. In particular we consider $bc$-ghost systems with
central charge $c=-2$ and study the replica formulation for
entanglement entropy for a single interval, and associated issues
arising in this case, notably pertaining to (i) the $SL(2)$ vacuum
coinciding with the ghost ground state, and (ii) the background charge
inherent in these systems which leads to particular forms for the
norms of states (involving zero modes). This eventually gives rise to
negative entanglement entropy. We also discuss a (logarithmic) CFT of
anti-commuting scalars, with similarities in some features. Finally we
discuss a simple toy model of two ``ghost-spins'' which mimics some of
these features.
\end{abstract}

\end{titlepage}

\newpage 
{\tiny %footnotesize
\begin{tableofcontents}
\end{tableofcontents}
}

%\vspace{5mm}

\section{Introduction}

$dS/CFT$ duality \cite{Strominger:2001pn,Witten:2001kn,Maldacena:2002vr} 
involves certain generalizations of gauge/gravity duality
\cite{Maldacena:1997re,Gubser:1998bc,Witten:1998qj,Aharony:1999ti}
conjecturing that de Sitter space is dual to a hypothetical Euclidean
non-unitary CFT that lives on the future boundary ${\cal I}^+$. More
precisely the late-time wavefunction of the universe $\Psi_{dS}$ with
appropriate boundary conditions is equated with the dual CFT partition
function $Z_{CFT}$ \cite{Maldacena:2002vr}. The dual CFT$_d$
energy-momentum tensor correlator $\langle T T\rangle$ obtained in a
semiclassical approximation $\Psi\sim e^{iS}$ exhibits central charge
coefficients ${\cal C}_d\sim i^{1-d}{R_{dS}^{d-1}\over G_{d+1}}$ in
$dS_{d+1}$. In $dS_4$, the central charge is real and negative: the 
dual CFT is thus reminiscent of ghost-like non-unitary theories. See \eg\ 
\cite{Strominger:2001gp,Balasubramanian:2001nb,Balasubramanian:2002zh,
Harlow:2011ke,Anninos:2011ui,Ng:2012xp,Das:2012dt,Anninos:2012ft,
Das:2013qea,Banerjee:2013mca,Das:2013mfa} for further work on 
$dS/CFT$: in particular in \cite{Anninos:2011ui}, a higher spin 
$dS_4$ duality was conjectured involving a 3-dim CFT of anti-commuting 
$Sp(N)$ (ghost) scalars.

Regardless of various details, it is perhaps of interest to better
organize our understanding of de Sitter space using this dual
conformal invariance, as well as constrain the properties such a
hypothetical CFT might have. In this context, it is perhaps of
interest to study entanglement entropy as a probe of $dS/CFT$. Some
attempts in this regard were begun in
\cite{Narayan:2015vda,Narayan:2015oka} by studying certain
generalizations of the Ryu-Takayanagi prescription
\cite{Ryu:2006bv,Ryu:2006ef} in $AdS/CFT$. We recall that in the
gravity approximation, the areas in Planck units of bulk minimal
surfaces (more generally extremal surfaces \cite{HRT}) anchored at the
subsystem interface and dipping into the bulk capture entanglement
entropy of the field theory subsystem (see \cite{HEEreview,HEEreview2}
for reviews). A different perspective on this appears more recently in
\cite{Lewkowycz:2013nqa} (see also
\cite{Hartman:2013mia,Faulkner:2013yia}).

In the de Sitter context, for strip-shaped subregions on a constant
Euclidean time slice of the future boundary, it was found in
\cite{Narayan:2015vda} that while real surfaces that might be of
relevance have vanishing area and are thus uninteresting, the areas of
certain codim-2 complex extremal surfaces have structural resemblance
with entanglement entropy of dual Euclidean CFTs (reviewed in
sec.~2). The coefficients of the leading divergent ``area law'' terms
in $dS_{d+1}$ resemble the central charges ${\cal C}_d\sim i^{1-d}
{R_{dS}^{d-1}\over G_{d+1}}$ of the CFT$_d$s appearing in the $\langle
T T\rangle$ correlators in \cite{Maldacena:2002vr}. In general the
areas thus obtained are negative or pure imaginary. In particular, in
$dS_4$, where the dual CFT central charge is $-{R_{dS}^2\over G_4}$~,
the area is negative.  (This leading divergence has also been studied
in \cite{Sato:2015tta}.)  In \cite{Narayan:2015oka}, this was
generalized to spherical subregions on a constant Euclidean time slice
of the future boundary.  In this case, in even boundary dimensions,
the area exhibits a subleading logarithmic divergence: the coefficient
of this piece matches precisely including numerical factors with the
coefficient of the subleading logarithmic divergence in the free
energy of the CFT on a sphere, using the $dS/CFT$ dictionary
$Z_{CFT}\equiv e^{-F} = \Psi_{dS}\sim e^{iS}$ with $\Psi_{dS}$ the
wavefunction of the universe in a classical approximation.  In
\cite{Das:2013mfa}, certain asymptotically de Sitter spaces were
argued to be gravity duals of the CFT at uniform energy density, and
thus are analogs of black branes: similar complex extremal surfaces in
these backgrounds exhibit a finite cutoff-independent extensive part
in the areas, analogous to thermal entropy, but again negative for the
$dS_4$ black brane \cite{Narayan:2015vda}. The resulting analysis
in all these cases ends up being equivalent to analytic continuation 
from the Ryu-Takayanagi expressions in $AdS/CFT$.

Towards regarding the areas of these complex extremal surfaces as
entanglement entropy, it is of interest to focus on the $dS_4$ case
where the central charge and areas are negative, and look for toy
models where negative entanglement entropy arises, if at all. It is
then natural to look to non-unitary Euclidean CFTs with negative
central charge, in particular in the context of 2-dim CFTs where
conformal symmetry is well-known to be especially powerful
\cite{Belavin:1984vu}. In this work, we revisit the replica
formulation \cite{Holzhey:1994we,Calabrese:2004eu,Calabrese:2009qy} of
entanglement entropy for a single interval, in certain 2-dim ghost
conformal field theories which have negative central charge as is
well-known. Focussing first on the $bc$-ghost system, sec.~3 (see \eg\
\cite{polchinskiTextBk,Blumenhagen:2013fgp}, as well as
\cite{Friedan:1985ge}, and more recently \cite{Saleur:1991hk,
  Kausch:1995py,Kausch:2000fu,Flohr:2001zs,Krohn:2002gh}), we note
that the $SL(2)$ vacuum is in general not equivalent to the ground
state: for the $c=-2$ $bc$-ghost system, the two coincide.  In this
case we find that the replica formulation is formally applicable in
the $SL(2)$ invariant vacuum, with the twist field operators
exhibiting negative conformal dimensions leading to negative
entanglement entropy (a study of a $\BZ_N$ orbifold of the $bc$-ghost
CFT corroborates the negative dimension of the twist field operators).
A crucial ingredient is that the ``inner product'' required for
nonvanishing observables reflects the charge asymmetry due to the
background charge inherent in these ghost systems: this is equivalent
to the presence of ghost zero mode insertions which soak up the
background charge. It is under these conditions that correlation
functions such as $\langle b c \rangle$ are nonvanishing. The replica
formulation then applies and leads to nonvanishing negative
entanglement entropy.

Then (in sec.~4) we discuss a Euclidean 2-dim conformal field theory
of complex anti-commuting ghost scalars $\chi,\ {\bar\chi}$. This is a
logarithmic CFT, discussed in \eg\ \cite{Gurarie:1993xq,Gurarie:1997dw,
Gurarie:2013tma,Kausch:1995py,Kausch:2000fu,Flohr:2001zs,Krohn:2002gh}.
This is in part motivated by the $Sp(N)$ higher spin 3-dim theory of
anti-commuting scalars conjectured to be dual to $dS_4$ in 
\cite{Anninos:2011ui} (studied previously in
\cite{LeClair:2006kb,LeClair:2007iy}).  Although one might expect
logarithms in correlation functions, the object of interest for a
single interval is a 2-point function of twist operators which
exhibits power law behaviour with no logarithms. This then gives the
same form as above for entanglement entropy. In sec.~5, we discuss a
toy model of two ``ghost spins'' with a non-positive inner product
which mimics some of the features of the ghost systems above: the
reduced density matrix obtained thus has some negative eigenvalues and
formally gives entanglement entropy for some states with negative real
part as well as an imaginary part. Sec.~6 has a discussion in part on
the negative areas of the complex $dS_4$ extremal surfaces and
negative EE, while Appendix A reviews some details on the replica
calculation.

It is important to note that the analysis here is in toy 2-dim ghost
CFTs with negative central charge and simply serves to illustrate that
negative EE can formally be obtained from appropriate generalizations
of the standard CFT replica technique. The resulting object, being
negative, does not satisfy properties of ordinary EE such as strong
subadditivity (Sec.~6). These 2-dim toy ghost CFTs here are motivated
by the 3-dim $Sp(N)$ higher spin theory in \cite{Anninos:2011ui} (and
are not to be considered as dual to $dS_3$ which has imaginary central
charge): it would be interesting to see if the present work can be
used to gain insights into the duals to $dS_4$ and $dS_3$.

\section{Reviewing de Sitter extremal surfaces}

Here we briefly review the study \cite{Narayan:2015vda,Narayan:2015oka} 
of bulk extremal surfaces anchored over strip- or sphere-shaped 
subregions on the future boundary ${\cal I}^+$ of de Sitter space 
$dS_{d+1}$ in the Poincare slicing or planar coordinate foliation. 
The metric is
\be\label{dSpoinc}
ds^2 = {R_{dS}^2\over\tau^2} (-d\tau^2 + dw^2 + d\sigma_{d-1}^2)\ ,
\ee
covering half of the spacetime, \eg\ the upper patch, with ${\cal I}^+$ 
at $\tau = 0$ and a coordinate horizon at $\tau=-\infty$.  This may be 
obtained by analytic continuation of a Poincare slicing of $AdS$, 
\be\label{AdStodS}
z \rightarrow -i\tau\ ,\qquad R_{AdS} \rightarrow -iR_{dS}\ ,\qquad 
t\ra -iw\ ,
\ee
where $w$ is akin to boundary Euclidean time, continued from time in 
$AdS$ (with $z$ the bulk coordinate).
The dual Euclidean CFT is taken as living on the future $\tau=0$
boundary ${\cal I}^+$.  We assume translation invariance with respect
to the boundary Euclidean time direction $w$, and consider a subregion
on a $w=const$ slice of ${\cal I}^+$. One might imagine that tracing
out the complement of this subregion then gives entropy in some sense
stemming from the information lost.  In the bulk, we mimic this by
appropriate de Sitter extremal surfaces on the $w=const$ slice,
analogous to the Ryu-Takayanagi prescription in $AdS/CFT$. 
Operationally these extremal surfaces begin at the interface of the 
subsystem (or subregion) and dip into the bulk time direction.

For a strip-shaped subregion on ${\cal I}^+$\ (with width say along 
$x$), parametrizing the spatial part in (\ref{dSpoinc}) as 
$d\sigma_{d-1}^2=\sum_{i=1}^{d-1} dx_i^2$ with $x\in \{x_i\}$, the 
$dS_{d+1}$ area functional on a $w=const$ slice is\ 
%\be\label{SdS-EEdS}
$S_{dS} = {R_{dS}^{d-1} V_{d-2}\over 4G_{d+1}} 
\int {d\tau\over\tau^{d-1}} \sqrt{({dx\over d\tau})^2-1 }$~.
After extremization this gives\ 
${\dot x}^2 = {-A^2\tau^{2d-2}\over 1-A^2\tau^{2d-2}}$ ,\ 
with ${dx\over d\tau} \equiv {\dot x}$, and $A^2$ is a conserved 
constant. Due to a crucial minus sign relative to the $AdS$ case, 
real surfaces (obtained with $A^2<0$) do not exhibit any turning point 
(where $|{\dot x}|\ra\infty$). Developing this further, it turns out 
that these give null surfaces with vanishing area, uninteresting from 
the point of view of entanglement entropy. On the other hand, it can 
be shown that certain complex extremal surfaces can be identified if 
the bulk time parameter takes an imaginary path $\tau=iT$. These are 
parametrized as $x(\tau)$, with\ $-{\dot x}^2=({dx\over dT})^2 
= {A^2 (-1)^{d-1} T^{2d-2}\over 1-(-1)^{d-1}A^2 T^{2d-2}}$~. These 
complex solutions to the extremization problem exist if the constant 
parameter $A^2$ satisfies $A^2>0$ for $dS_4$ (and odd $d$ more 
generally), and $A^2<0$ for $dS_3, dS_5$ (and even $d$). These are 
smooth surfaces exhibiting a turning point $T_*^{d-1}={1\over A}$ 
where\ $|{dx\over dT}|\ra\infty$ .\ Thus for a strip subregion, the 
area has the form
\be\label{EEdSd+1}
S_{dS} = -i {R_{dS}^{d-1}\over 4G_{d+1}} V_{d-2} 
\int_{\tau_{UV}}^{\tau_*} {d\tau\over\tau^{d-1}} 
{1\over\sqrt{1-(-1)^{d-1}A^2\tau^{2d-2}}}\ =\
i^{1-d} {R_{dS}^{d-1}\over 2G_{d+1}} V_{d-2} 
 \Big({1\over\epsilon^{d-2}} - c_d {1\over l^{d-2}}\Big) ,\  
\ee
where $\tau_{UV}=i\epsilon$ and $\tau_*=il$, and the integral ends up
being similar to that in $AdS$ (with corresponding constant $c_d$). 
The area of these surfaces passes several checks from the point of 
view of regarding these as entanglement entropy in the dual Euclidean 
CFT in a $dS/CFT$ perspective.
$S_{dS}$ in (\ref{EEdSd+1}) bears structural resemblance to 
entanglement entropy in a dual CFT with central charge 
${\cal C}_d\sim i^{1-d} {R_{dS}^{d-1}\over G_{d+1}}$.
The first term $S_{dS}^{div}\sim i^{1-d} {R_{dS}^{d-1}\over G_{d+1}} 
{V_{d-2}\over\epsilon^{d-2}}$ resembles an area law divergence 
\cite{AreaLaw1,AreaLaw2}, proportional to the area of the interface
between the subregion and the environment, in units of the ultraviolet
cutoff. It appears independent of the shape of the subregion,
expanding the area functional and assuming that ${\dot x}$ is small 
near the boundary $\tau_{UV}$. Rewriting this as
${\cal C}_d {V_{d-2}\over\epsilon^{d-2}}$, we see that it is also 
proportional to the central charge 
${\cal C}_d\sim i^{1-d} {R_{dS}^{d-1}\over G_{d+1}}$ representing 
in some sense the number of degrees of freedom in the dual 
(non-unitary) CFT: these arose in the $\langle T T\rangle$ 
correlators obtained in \cite{Maldacena:2002vr}. In $dS_4$, the 
central charge ${\cal C}\sim -{R_{dS}^2\over G_4}$ is real and 
negative, while in $dS_3, dS_5$, it is imaginary. The second term 
is a finite cutoff-independent piece.

Likewise for a spherical subregion on a constant boundary Euclidean
time slice $w=const$, there are complex extremal surfaces of the form
$r(\tau) = \sqrt{l^2 + \tau^2}$ with $\tau=iT$. (There are also real 
null surfaces with vanishing area, as well as those of the form above, 
for $\tau$ real, with no finite cutoff-independent pieces.) 
Then for even boundary dimensions $d$, the area of these complex 
extremal surfaces exhibits a subleading logarithmic divergence
whose coefficient is related to the trace anomaly of the dual
Euclidean CFT$_d$. This can be seen to match precisely with the
coefficient of a corresponding logarithmic divergence in the free
energy of the CFT on a sphere, obtained via the $dS/CFT$ dictionary
with the wavefunction of the universe in a classical approximation 
$Z_{CFT}=e^{-F}=\Psi_{dS}\sim e^{iS_{cl}}$.

The resulting expressions obtained in the end amount to analytic
continuation from the Ryu-Takayanagi expressions
\cite{Ryu:2006bv,Ryu:2006ef} for holographic entanglement entropy in
$AdS/CFT$, although this was not obvious to begin with starting
directly in de Sitter space.

\vspace{1mm}

\underline{{\bf 4-dim de Sitter space, $dS_4$:}}\ \ For a strip 
subregion of width $l$ on the future boundary ${\cal I}^+$ of de 
Sitter space $dS_4$ restricting to a constant boundary Euclidean time 
slice, the complex extremal surfaces have area
\be\label{EEdS4strip}
S_{dS} \sim\ -{R_{dS}^2\over G_4} \Big( {V_1\over\epsilon} - {V_1\over l} \Big)
\ee
following from (\ref{EEdSd+1}), while for a spherical subregion of 
radius $l$, we have
$S_{dS} = -{\pi R_{dS}^2\over 2G_4} ({l\over\epsilon} - 1)$~.\
The finite constant cutoff-independent piece ${\pi R_{dS}^2\over 2G_4}$ 
is a universal term, positive definite (note that it resembles de 
Sitter entropy).
Thus for $dS_4$, we see that the areas of these complex extremal
surfaces are real and negative.  Structurally they resemble
entanglement entropy in a 3d CFT with negative central charge.
%\be\label{EEdS4sph}%\ee %{R_{dS}^2\Omega_1\over 4G_4} (-il)
%\int_{\tau_{UV}}^{\tau_*} {d\tau\over\tau^2}\ 

Although we have obtained complex surfaces as solutions to this 
extremization question, the resulting expressions are all real: 
specifically with $\tau=iT$, we rewrite the expressions in terms of 
$T=|\tau|$, obtaining
\bea
&& {\Delta x\over 2} = {l\over 2}\ =\ \int_0^{T_*} {(T^2/T_*^2)\ dT\over 
\sqrt{1-(T^4/T_*^4)}}\ \equiv\ \int_0^{|\tau_*|} {({|\tau|^2\over |\tau_*|^2}) 
\ d|\tau|\over \sqrt{1-{|\tau|^4\over |\tau_*|^4}}}\ ,\nonumber\\
&& S_{dS_4}\ =\ -i {R_{dS}^2\over 4G_4} V_1 \int_{\tau_{UV}}^{\tau_*} 
{d\tau\over\tau^2} {1\over\sqrt{1-\tau^4/\tau_*^4}}\ =\ 
-{R_{dS}^2\over 4G_4} V_1 \int_{|\tau_{UV}|}^{|\tau_*|} 
{d|\tau|\over|\tau|^2} {1\over\sqrt{1-{|\tau|^4\over|\tau_*|^4}}}\ .
\eea
These resulting expressions are of course related in a very simple way
to the Ryu-Takayanagi $AdS$ expressions. The point of this rewriting
is to make manifest the mapping from the complex $\tau$-path to
corresponding real $\tau$-values in $dS_4$.  For each $\tau=iT$
ranging from $\tau_{UV}$ to $\tau_*$, we have a corresponding real
$\tau$ given simply by $\tau_R\equiv -T=-|\tau|$, ranging over
$(-\epsilon, -T_*)$ in the bulk de Sitter space. The strip width $l$
is then related to this bulk $dS_4$ time $\tau_R$ as\ $l\sim T_*$:
thus increasing $l$ means larger $|\tau_R|$, \ie\ further back in the
past.  These expressions while real-valued are of course completely
different from the extremization process restricting to real
$\tau$-values in $dS_4$ \cite{Narayan:2015vda}: the latter lead to
null surfaces with vanishing area and thus no bearing on entanglement
entropy.

The nonunitary dual CFTs in the de Sitter context, 
with negative or imaginary central charge, contain
operators with complex conformal dimensions in general. For instance, 
bulk scalar modes of mass $m$ in $dS_{d+1}$ obey the wave equation\
${1\over\sqrt{-g}}\del_\mu (g^{\mu\nu}\sqrt{-g}\del_\nu\phi) - m^2\phi = 0$. 
This becomes\ $\tau^2{\ddot\varphi} - (d-1)\tau{\dot\varphi} 
+ (k^2\tau^2 + m^2R_{dS}^2)\varphi = 0$\ 
for modes of the form $\varphi(\tau) e^{ik_ix^i}$.
Near the future boundary $\tau\ra 0$, these are approximated as 
$\varphi \sim\tau^\Delta$, giving the dual conformal dimensions 
$\Delta(\Delta-d)=-m^2R_{dS}^2$ \ie\ 
$\Delta = {d\over 2}\pm\sqrt{{d^2\over 4}-m^2R_{dS}^2}$ (analogous 
to $AdS/CFT$).
So for bulk de Sitter modes with sufficiently large mass, the dual 
operator conformal dimensions $\Delta$ are complex:\
$\Delta = {d\over 2}\pm i\sqrt{m^2R_{dS}^2-{d^2\over 4}}\ \sim\
{d\over 2}\pm i mR_{dS}$ for\ $mR_{dS}\gg 1$.\
This is a feature of $dS/CFT$ and more generally of field asymptotics
in de Sitter space\ (in the higher spin $dS_4/CFT_3$ case 
\cite{Anninos:2011ui}, the conformal dimensions are real-valued). 
However we note that for all $mR_{dS}\gtrsim {d\over 2}$, the real 
part of the conformal dimensions is fixed, and positive with 
Re$\Delta = {d\over 2}$~.
For bulk modes with $mR_{dS}\leq {d\over 2}$, the 
conformal dimensions are real and positive. Thus in the 
approximation of bulk de Sitter theories consistently truncated to 
only modes with masses satisfying $mR_{dS}\leq {d\over 2}$~, all 
dual conformal dimensions are real and with $O(1)$ values. 

It is worth noting that these complex solutions lie outside the
original de Sitter coordinate range (where $\tau$ is real) so one may
conclude that, strictly speaking, there are no extremal surfaces with
nonvanishing area, and correspondingly no notion of dual entanglement
entropy in de Sitter space in the sense here, based on the
Ryu-Takayanagi formulation (also the dual theory is Euclidean so the
boundary Euclidean time slice used here may be regarded as ad hoc). If
instead we consider the complex extremal surfaces, focussing on
$dS_4$, we have seen negative areas (and thereby negative EE).
Towards understanding this from a CFT point of view, we will study
some toy models of 2-dim CFTs with negative central charge and
entanglement entropy from a replica formulation. Specifically we focus
on some 2-dim ghost CFTs, in particular the $bc$-ghost system and the
CFT of anti-commuting ghost scalars.

\section{$bc$-ghost CFTs}

The $bc$-ghost CFT is familiar from worldsheet string theory: for our 
present purposes, see \eg\ \cite{polchinskiTextBk,Blumenhagen:2013fgp},
as well as \cite{Friedan:1985ge}, and more recently 
\cite{Saleur:1991hk,Kausch:1995py,Kausch:2000fu,Flohr:2001zs,Krohn:2002gh}. 
Our discussion here is to some extent simply a review of some relevant 
aspects, but adapted to the present context. 
The action and conformal weights 
\be
S={1\over 2\pi} \int d^2z~b{\bar\del}c\ , \qquad\qquad\qquad
h_b=\lambda\ ,\qquad h_c=1-\lambda\ ,
\ee
for any $\lambda$ ensure conformal invariance in this holomorphic 
part of the CFT. The conformal fields $b, c$, are anticommuting 
variables. (We will suppress writing the anti-holomorphic parts in most 
of our discussion.)\ We have the equations of motion\ 
${\bar\del}c(z) = 0 = {\bar\del}b(z)$, and 
${\bar\del}b(z) c(w) = 2\pi \delta^2(z-w, {\bar z}-{\bar w})$. This 
gives the singular parts of the OPEs
\be
b(z) c(w) \sim {1\over z-w}\ , \qquad  c(z) b(w) \sim {1\over z-w}\ , 
\ee
and the energy-momentum tensor and central charge
\bea
T(z) &=& :(\del b)c: - \lambda \del (:bc:)\
=\ {1\over 2} \big( :(\del b)c: - :b\del c: \big)
+ {1\over 2} Q \del ( :bc: )\ ,\\
&& \qquad\ 
c=1-3(2\lambda-1)^2 = 1-3Q^2\ ,\qquad\ Q=1-2\lambda\ .
\eea
The central charge can be seen from the $TT$ OPE which is of standard 
form.\ $Q$ is the background charge for the ghost system and is 
nonvanishing except when $\lambda={1\over 2}$ (which corresponds to 
two $c={1\over 2}$ free fermions).\ The mode expansions can be written as
\be\label{bcmodeExp}
b(z) = \sum_{m\in \BZ} {b_m\over z^{m+\lambda}}\ ,\qquad
c(z) = \sum_{m\in \BZ} {c_m\over z^{m+1-\lambda}}\ .
\ee
Inverting we have\ $b_m = \oint {dz\over 2\pi i} z^{m+\lambda-1} b(z)\ ,\
c_m = \oint {dz\over 2\pi i} z^{m-\lambda} c(z)$.\
These and the OPEs lead via the contour arguments to the oscillator algebra
and Virasoro generators
\bea\label{LmL0}
&& \qquad\qquad \{ b_m, c_n\} = \delta_{m+n, 0}\ ,\qquad 
\{ b_m, b_n\} = 0\ , \qquad \{ c_m, c_n\} = 0\ ,\nonumber\\
&& L_m = \sum_{n=-\infty}^\infty (m\lambda - n) b_n c_{m-n}\quad
[m\neq 0]\ ;\qquad
L_0 = \sum_{n>0} n (b_{-n}c_n + c_{-n}b_n) + {\lambda (1-\lambda)\over 2}\ ,
\qquad
\eea
which satisfy the Virasoro algebra
$[L_m, L_n]=(m-n) L_{m+n} + A(m) \delta_{m+n,0}$, where $A(m)$ is a 
central term. Since $b, c$ are real fields, the hermiticity relations are 
$b_m^\dag=b_{-m},\ c_m^\dag=c_{-m}$, and $L_m^\dag=L_{-m} ,\ L_0^\dag=L_0$.
The zero mode sector $\{b_0, c_0\}=1$ gives two states 
$|\downarrow\rangle,\ |\uparrow\rangle$, with
\be\label{ghostgndst}
b_0|\downarrow\rangle=0 ,\quad c_0|\downarrow\rangle=|\uparrow\rangle\ ,
\qquad b_0|\uparrow\rangle=|\downarrow\rangle ,\quad c_0|\uparrow\rangle = 0\ ,
\ee
and all $b_n, c_n,\ n>0$ annihilating both 
$|\downarrow\rangle,\ |\uparrow\rangle$. It is conventional to group 
$b_0$ with the annihilation operators and take $|\downarrow\rangle$ as 
the ghost ground state.

%The state-operator correspondence and the mode expansions above can be
We define the $SL(2,\BZ)$ invariant vacuum $|0\rangle$ as the vacuum 
where the energy-momentum tensor and the conformal fields are regular 
at the origin $z=0$ of the $z$-plane:
\be\label{SL2vacLm}
|0\rangle:\qquad\qquad 
T(z)|0\rangle = \sum_m {L_m\over z^{m+2}} |0\rangle = regular \quad 
\Rightarrow\quad L_m |0\rangle = 0\ ,\quad m\geq -1\ .
\ee
Thus $L_0, L_{\pm 1}$ annihilate $|0\rangle$. In particular 
$L_0|0\rangle = 0$ implies that the $SL(2)$ vacuum has zero $L_0$ 
eigenvalue. Similarly, using the mode expansions (\ref{bcmodeExp}) 
above, we see that regularity of $b(z)|0\rangle$ and $c(z)|0\rangle$ 
at $z=0$ gives
\be\label{SL2vacbc}
|0\rangle:\qquad\qquad b_{m\geq 1-\lambda} |0\rangle = 0\ ,\quad
c_{m\geq \lambda} |0\rangle = 0\ .
\ee

The ghost ground state, of lowest $L_0$ eigenvalue, does not necessarily 
coincide with the $SL(2)$ vacuum $|0\rangle$ however, due to the 
shifts in the mode expansions (\ref{bcmodeExp}). In particular,\ 
$L_0|\downarrow\rangle = {\lambda (1-\lambda)\over 2} |\downarrow\rangle$
so that for $\lambda>1$, the $L_0$ eigenvalue of $|\downarrow\rangle$ 
is negative.

The $U(1)$ charge symmetry $\delta b=-i\epsilon b,\ \delta c=i\epsilon c$ 
gives the ghost current\ $j(z)=-:bc:$,\ with the OPEs\ \
$j(z)b(w) \sim -{1\over z-w} b(w),\ \ j(z)c(w) \sim {1\over z-w} c(w)$,\ \
and\ \ $j(z) j(w) \sim {1\over (z-w)^2}$.\ \
The $Tj$ OPE exhibits an anomalous transformation 
\be\label{TjOPE}
T(z) j(w) \sim 
{Q\over (z-w)^3} + {1\over (z-w)^2} j(w) + {1\over z-w} \del j(w)\ ,
\ee
the non-tensor term in the OPE arising from the background charge 
$Q=1-2\lambda$. This leads to a finite transformation\ 
$(\del_zw) j'(w) = j(z) - {Q\over 2} {\del_z^2 w\over \del_zw}$.\ 
The ghost number $N_g$ on the cylinder and that on the plane $N_g^z$ are
\be\label{NgNgz}
N_g = \int_0^{2\pi} {dw\over 2\pi i}~ j^{cyl}(w) 
= \sum_{n=1}^\infty (c_{-n}b_n - b_{-n}c_n) + c_0b_0 - {1\over 2} ,
\qquad N_g^z = \oint {dz\over 2\pi i} j(z) = N_g - {Q\over 2} .
\ee
The ghost number on the plane $N_g^z$, counting $j_0$ charge, is 
obtained from the finite transformation above for the cylinder coordinate 
$w=\log z$, the anomalous term leading to the shift.  
%$-{Q\over 2} {-1/z^2\over 1/z}$.
We see that $[N_g, b_m]=-b_m,\ [N_g, c_m]=c_m$. Conventionally, the 
ghost number of states is taken as
$N_g|\downarrow\rangle = -{1\over 2} |\downarrow\rangle,\ 
N_g|\uparrow\rangle = {1\over 2} |\uparrow\rangle$.\ This gives 
$N_g^z|\downarrow\rangle = -{Q+1\over 2} |\downarrow\rangle 
= (\lambda-1)|\downarrow\rangle$, and implies that $N_g^z |0\rangle = 0$.
Since the $SL(2)$ vacuum $|0\rangle$ satisfies (\ref{SL2vacLm}), 
(\ref{SL2vacbc}), the $b_m$ operators for $1-\lambda \leq m < 0$ act 
as annihilation operators on $|0\rangle$ although they are creation 
operators for $|\downarrow\rangle$. This implies that the $SL(2)$ 
vacuum interpreted as a Fermi sea is filled for the 
$b_{1-\lambda \leq m < 0}$ oscillators, and we can identify this as
\be\label{SL2Fermisea}
|0\rangle = b_{-1} b_{-2} \ldots b_{1-\lambda} |\downarrow\rangle = 
\prod_{1-\lambda \leq m < 0} b_m |\downarrow\rangle\ \qquad\qquad 
[\lambda \neq 1]\ .
\ee
This then implies that 
\be\label{SL2NgNgz}
N_g |0\rangle = \Big(-{1\over 2} - (\lambda - 1)\Big) |0\rangle = 
{Q\over 2} |0\rangle\ , \qquad\ \  \ie \qquad\ \  N_g^z |0\rangle = 0\ ,
\ee
since the $b$-oscillators cancel the $N_g^z$ ghost charge $\lambda-1$ 
of $|\downarrow\rangle$. In other words, the $SL(2)$ vacuum $|0\rangle$ 
has zero $j_0$ charge on the plane.

\vspace{1mm}

The presence of the background charge $Q$ forces a particular form for 
the adjoints of states and correspondingly their norms. 
The OPE (\ref{TjOPE}) and\ $j(z)=\sum_m {j_m\over z^{m+1}}$\ gives
\be
[L_m, j_n] = -nj_{m+n} + {1\over 2} Q m(m+1) \delta_{m,-n}\ .
\ee
In particular, while $j_n^\dag=-j_n$ for $n\neq 0$ (with 
$j_n=-\sum_m b_mc_{n-m}$),\ the asymmetry between $m=+1$ and $m=0,-1$, 
above gives
\be
[L_1, j_{-1}]=j_0+Q\ ,\quad [L_1, j_{-1}]^\dag = [L_{-1},j_1] = -j_0\ 
\quad \Rightarrow\quad  j_0^\dag = -(j_0+Q)\ ,
\ee
so that there is an asymmetry in the $bc$-system from the charge operator 
$j_0$, indicating the presence of the background charge $Q$. In 
particular charge neutral operators have vanishing correlation functions: 
with $[j_0,O_p]=pO_p$, and $j_0|q\rangle = q|q\rangle$, we have
\be\label{pqq'Q}
p\langle q'|O_p|q\rangle = \langle q'|[j_0,O_p]|q\rangle 
= (-q'-Q-q) \langle q'|O_p|q\rangle\ ,
\ee
so that the correlation function is non-vanishing only if\ $p=-(q+q'+Q)$. 
Considering $p=0$, we have $q'=-q-Q$ for a nonvanishing correlation 
function, so that the inner products of the states $|q\rangle$ can be 
normalized as
\be\label{qnorm}
\langle -q-Q|q\rangle = 1\ .
\ee
Since nonzero correlation functions only arise after inserting
operators that soak up the background charge $Q$, this normalization
can be interpreted as the inner product of the $|q\rangle$ in-state with
an out-state $\langle -q-Q|$ which is the adjoint of $|q\rangle$. In
particular for the $SL(2)$ vacuum $|0\rangle$ taken as the in-state,
we require the out-state $\langle -Q|$ for a nonvanishing correlator. 
Recalling that the difference in the number of zero modes is\
$\#_0^c - \#_0^b = -{1\over 2} Q\chi = -Q$ on the sphere (with Euler
number $\chi=2$), we see that the normalization (\ref{qnorm})
precisely corresponds to inserting the appropriate number of operators
that soak up the zero modes of the $bc$-ghost system in order to
obtain nonvanishing correlation functions.
From a path integral point of view, we see that\
the vacuum partition function $\langle 0| 0\rangle \equiv 
\int Db Dc\ e^{-S}$ vanishes due to Grassmann integration over unbalanced 
zero modes of the $b,c$-fields (which do not appear in the action). 
The requisite number of zero mode insertions makes this nonvanishing and 
amounts to the normalization (\ref{qnorm}).

%\vspace{1mm}

To illustrate this, consider the $\lambda=2$ $bc$-system, with 
background charge $Q=-3$ and central charge $c=-26$, which arises in 
the reparametrization ghosts of the string worldsheet theory. The 
$(b, c)$-fields have conformal weights $(h_b, h_c)=(2,-1)$. The 
$SL(2)$ invariant vacuum $|0\rangle$ satisfies
$b_{m\geq -1} |0\rangle = 0 ,\ c_{m\geq 2} |0\rangle = 0$,\
and we identify $|0\rangle \equiv b_{-1}|\downarrow\rangle$. 
%or $|\downarrow\rangle = c_1 |0\rangle$.\ 
In this case, we have\ 
$L_0 |\downarrow\rangle = - |\downarrow\rangle$.\
We see that $\langle \downarrow| \downarrow\rangle = 
\langle \downarrow| \{ b_0, c_0 \} |\downarrow\rangle = 0$ since 
$b_0$ and $b_0^\dag=b_0$ annihilate the states. Since $Q=-3$, the 
normalization (\ref{qnorm}) becomes\ $\langle +3| 0 \rangle = 1$ for 
the $SL(2)$ vacuum. The difference in zero modes is\ 
$\#_0^c - \#_0^b = -{1\over 2} Q\chi = 3$ on the sphere, and the 
smallest nonvanishing correlation function (which is not annihilated 
by inserting the oscillator relation) requires the insertion of three 
$c$-operators, \eg\ $\langle 0| c_{-1} c_0 c_1 (\ldots) |0\rangle$,\
as is well-known from worldsheet string amplitudes.

\subsection{$bc$-ghosts with $\lambda=1$}

In this case, the conformal weights of the $(b, c)$-fields, the 
background charge $Q$ and the central charge $c$ are
\be\label{lambda=1Qc}
\lambda=1:\qquad\qquad (h_b, h_c)=(1,0)\ ,\qquad\ \ Q=-1\ ,\qquad\ \ c=-2\ .
\ee
The mode expansions (\ref{bcmodeExp}) in this case are
\be\label{lambda=1bcExp}
b(z) = \sum_m {b_m\over z^{m+1}}\ ,\qquad\qquad
c(z) = \sum_m {c_m\over z^m}\ .
\ee
From (\ref{SL2vacbc}), the $SL(2)$ invariant vacuum satisfies
\be\label{lambda=1SL2vacbc}
|0\rangle:\qquad\qquad b_{m\geq 0} |0\rangle = 0\ ,\qquad\ \ 
c_{m\geq 1} |0\rangle = 0\ ,
\ee
which coincide with the conditions (\ref{ghostgndst}) defining the 
ghost ground state $|\downarrow\rangle$. Thus instead of 
(\ref{SL2Fermisea}), we identify
\be
\lambda=1:\qquad\qquad |0\rangle = |\downarrow\rangle\ ,
\ee
\ie\ the $SL(2)$ invariant vacuum $|0\rangle$ coincides 
with the ghost ground state $|\downarrow\rangle$. To see 
consistency of this, we note from (\ref{SL2NgNgz}) that\ 
$N_g|0\rangle = -{1\over 2} |0\rangle$, which coincides with the 
ghost number of $|\downarrow\rangle$: also (\ref{NgNgz}) gives
$N_g^z|\downarrow\rangle = (\lambda-1)|\downarrow\rangle = 0$.\ 
The Virasoro generators (\ref{LmL0}) become
\be
L_m = \sum_{n=-\infty}^\infty (m - n) b_n c_{m-n}\qquad
[m\neq 0]\ ; \qquad\ \ \
L_0 = \sum_{n>0} n (b_{-n}c_n + c_{-n}b_n)\ ,
\ee
so that $L_0|\downarrow\rangle = L_0|0\rangle = 0$.\ Thus in this case, 
the $SL(2)$ vacuum is also the ground state of the system with $L_0=0$.

The field of lowest conformal dimension is $c(z)$ with $h_c=0$. 
However since this is a first order system of anticommuting fields, 
the $b, c$, OPEs are not logarithmic, with the $cc$ OPE vanishing.
$c_0$ does not annihilate $|0\rangle$: instead 
$c_0|0\rangle = |\uparrow\rangle$. The state $c_0|0\rangle$ with 
$L_0=0$ is degenerate with $|0\rangle$. The ghost numbers are
$N_gc_0|0\rangle = {1\over 2}c_0|0\rangle$ and 
$N_g^zc_0|0\rangle=c_0|0\rangle$.

The background charge here is $Q=-1$ and the normalization 
(\ref{qnorm}) in this case becomes\ 
$\langle -q+1|q\rangle = 1$: for the $SL(2)$ vacuum, this is 
$\langle +1|0\rangle = 1$. Using (\ref{lambda=1bcExp}) and 
(\ref{lambda=1SL2vacbc}), the smallest nonvanishing correlation 
function is
\be
\langle 0| c(z) |0\rangle = \langle 0| \sum_m {c_m\over z^m} |0\rangle 
= \langle 0| c_0 |0\rangle\ \equiv\ \langle +1|0\rangle\ ;
\qquad\quad \langle +1| = \langle 0|c_0\ ,
\ee
so that the adjoint of the $SL(2)$ vacuum is the state 
$\langle 0|^\dag = \langle 0|c_0$, with $c_0^\dag = c_0$.
More general nonvanishing correlation functions in the $|0\rangle$ 
in-state are obtained as above, with the out-state $\langle 0|c_0$. 
For instance
\be
\langle b(z) c(w) \rangle_0 \equiv \langle 0|c_0\ \sum_{m, n} 
{b_m\over z^{m+1}} {c_n\over w^n} |0\rangle = \langle 0|c_0\ %{1\over z} 
\sum_{m=0}^\infty
{w^m\over z^{m+1}} b_m c_{-m} |0\rangle = {1\over z} {1\over 1-{w\over z}} 
\langle 0|c_0 |0\rangle\ = {1\over z-w}\ ,
\ee
which is the expected form of the 2-point function. Note that without 
the $c_0$-insertion, we have\
$\langle 0| b(z) c(w) |0\rangle = {1\over z-w} \langle 0| 0\rangle = 0$, 
using \eg\ arguments such as (\ref{pqq'Q}). Thus the short distance 
structure of the $bc$-ghost system is only visible in correlation 
functions with insertions that appropriately cancel the background charge.

With the norms of states defined in this manner, we see that there 
are various negative norm states in this system, as expected from the 
negative central charge. Among the simplest is \eg\ 
$(b_{-1}-c_{-1})|0\rangle$ with adjoint $\langle 0|c_0 (b_1 - c_1)$: 
this has norm\
\be
\langle 0| c_0 (b_1 - c_1) (b_{-1}-c_{-1})|0\rangle 
= -\langle 0|c_0 \big(\{b_1, c_{-1}\} + \{c_1, b_{-1}\}\big) |0\rangle 
= -2\langle 0|c_0|0\rangle = -2\ .
\ee
Relatedly, the state $(b_{-1}+c_{-1})|0\rangle$ has positive norm, while 
the states $b_{-1}|0\rangle,\ c_{-1}|0\rangle$, have zero norm.
There is a plethora of negative norm states as we go to higher levels, 
built with oscillators as above.

\subsection{The replica calculation and entanglement entropy}

Revisiting the replica formulation of Calabrese, Cardy 
\cite{Calabrese:2004eu,Calabrese:2009qy}, the calculation of 
entanglement entropy for a subsystem $A$ consisting of a single 
interval is obtained as\ 
$S_{EE}^A = -\displaystyle{\lim_{n\to 1}}\ \del_n tr\rho_A^n$\ 
where $tr\rho_A^n$, is the path integral in the replica space 
consisting of $n$-copies of the original space, with adjacent copies 
appropriately sewn together, in the presence of a cut representing 
the interval. This partition function itself can be expressed in 
terms of a certain 2-point correlation function after a conformal 
transformation.

In more detail (see Appendix A), consider a 2-dim Euclidean CFT with 
central charge $c$, which we imagine to be a ghost CFT with negative 
central charge $c<0$ of the sort above. Let us use complex coordinates 
$w=x+it_E$ and ${\bar w}$, where $t_E$ is Euclidean time. 
From the point of view of regarding this as an intrinsically Euclidean
theory, we are assuming translation invariance along one direction,
which we take as Euclidean time: this allows us to decompose the
Euclidean CFT into Euclidean time slices.
Time evolution pertains to this Euclidean time and so does the
entanglement entropy we are discussing for the subsystem in question,
which is a spatial interval on a constant Euclidean time slice. So 
consider a subsystem $A$ defined by a single interval stretched 
between $x=u$ and $x=v>u$ on a fixed time slice $t_E=const$.\
Under a conformal transformation $w\ra z$, the energy-momentum tensor 
transforms as
\be\label{TzConfTransf}
T(w) = (\del_w z)^2 T(z) + {c\over 12} \{z,w\}\ ,
\ee
where\ $\{z,w\} = {2\del_w^3z \del_wz - 3 (\del_w^2z)^2\over 2(\del_wz)^2}$\ 
is the Schwarzian derivative.\ 
The replica $w$-space is transformed into the the $z$-plane by the 
conformal transformation (\ref{zwConftrans}) given by\ 
$z = ({w-u\over w-v})^{1/n}$\ (see Appendix A). Assuming that on the 
$z$-plane there are no insertions of any nontrivial operators gives
\be\label{Tz=0}
\langle T(z)\rangle_{\BC} = 0\ ,
\ee
which is equivalent to taking the $z$-plane to represent the CFT 
ground state. Now taking expectation values (with ${\cal R}_n$ the 
$n$-sheeted $w$-space), we have 
\be\label{T(w)vev}
\langle T(w)\rangle_{{\cal R}_n} = {c\over 12} \{z,w\} 
= {c (1-{1\over n^2})\over 24} \ {(v-u)^2\over (w-u)^2 (w-v)^2}\ 
=\ {\langle T(w) \Phi_n(u) \Phi_{-n}(v) \rangle \over 
\langle \Phi_n(u) \Phi_{-n}(v) \rangle}\ .
\ee
The boundary conditions at $u, v$ are equivalent to the insertion of 
twist operators $\Phi_n(u), \Phi_{-n}(v)$ at $w=u, v$ respectively. 
Comparing the last expression with the standard form for the 
$\langle T(z) O_1(z_1) O_2(z_2)\rangle$ 3-point function in CFT 
(see Appendix A) which is related to the 2-point function 
$\langle O_1(z_1) O_2(z_2)\rangle$, the conformal dimensions of the 
twist operators can be read off.

For the $bc$-ghost CFTs in question here, the formulation above must 
be refined further.

\vspace{1mm}

\noindent (1)\ Firstly, the condition (\ref{Tz=0}) is equivalent to
the statement that the $z$-plane represents the $SL(2)$ invariant
vacuum $|0\rangle$ rather than the ghost ground state
$|\downarrow\rangle$.  It is the $SL(2)$ vacuum (\ref{SL2vacLm}) which
is defined by regularity of the energy-momentum tensor, and which
naturally enters the condition (\ref{Tz=0}) defining the $z$-plane
after the conformal transformation (\ref{zwConftrans}), and where the
resulting correlation functions can be studied. Equivalently the
state-operator correspondence maps the identity operator to the
$SL(2)$ vacuum through (\ref{SL2vacLm}).

In general, the $SL(2)$ vacuum with $L_0=0$ is not the ghost ground
state $|\downarrow\rangle$ which has\
$L_0|\downarrow\rangle={\lambda(1-\lambda)\over 2}|\downarrow\rangle$,
the $L_0$-eigenvalue being negative for $\lambda>1$. In the state
$|\downarrow\rangle$, naively the condition (\ref{Tz=0}) is not
well-defined near $z=0$, due to \eg\ contributions such as\ ${L_0\over
  z^2}|\downarrow\rangle$. Thus while the replica calculation in the
$SL(2)$ vacuum appears to be formally valid, the interpretation is not
entirely clear: the $SL(2)$ vacuum is perhaps best regarded as an
excited state consisting of a partially filled Fermi sea
(\ref{SL2Fermisea}), leaving the question of what happens in the
ground state $|\downarrow\rangle$.

\vspace{1mm}

\noindent (2)\ Secondly, note that of course the condition 
$\langle T(z)\rangle_{\BC} = 0$ is vacuous unless we use the 
normalizations of states given by (\ref{qnorm}). For instance, taking 
$\langle 0| T(z)|0\rangle$ instead gives a trivial condition since 
$\langle 0| 0\rangle = 0$: the background charge inherent in the 
$bc$-ghost system is not cancelled without appropriate ghost zero 
mode insertions, equivalent to the normalizations (\ref{qnorm}).
Thus (\ref{Tz=0}) using the $SL(2)$ vacuum must be regarded as
\be
\langle -Q|\ T(z)\ |0\rangle = 0\ ,
\ee
where $Q$ is the background charge of the $bc$-ghost system, in 
accord with (\ref{qnorm}).\\

For the case $\lambda=1$ corresponding to the $bc$-system with $c=-2$,
the $SL(2)$ vacuum is the ghost ground state, and the state of lowest
$L_0$ eigenvalue. Its adjoint is the state $\langle 0|c_0 = 
\langle\uparrow|$ which also has $L_0=0$.  Although there is a
plethora of negative norm states (norms defined via (\ref{qnorm})),
the $L_0$ eigenvalues are positive. Thus the replica formulation in
the $SL(2)$ vacuum can be taken to represent the entanglement entropy 
in the ground state.

\vspace{2mm}

The $n$-sheeted replica theory is defined by boundary conditions 
\bea\label{replicabc}
b_k(e^{2\pi i} (w-u)) = b_{k+1}(w-u)\ , &&
c_k(e^{2\pi i} (w-u)) = c_{k+1}(w-u)\ ,\nonumber\\
b_k(e^{2\pi i} (w-v)) = b_{k-1}(w-v)\ , && 
c_k(e^{2\pi i} (w-v)) = c_{k-1}(w-v)\ ,
\qquad  k=1,\ldots,n , \qquad
\eea
so that\ $(b,c)_k\ra (b,c)_{k+1}$ going around $w=u$ as 
$w-u\ra e^{2\pi i} (w-u)$\ and\ $(b,c)_k\ra (b,c)_{k-1}$ under 
$w-v\ra e^{2\pi i} (w-v)$,\ with the $n$-th sheet connecting back to 
the first. If one encircles both $w=u$ and $w=v$, the boundary 
conditions are trivial, \ie\ far from the interval the replica space 
is essentially a single sheet.
The fields and boundary conditions above can be diagonalized by defining\
\be\label{replicabcdiag}
{\tilde b}_k = {1\over n} \sum_{l=1}^n e^{2\pi ilk/n} b_k\ ,\qquad
{\tilde c}_k = {1\over n} \sum_{l=1}^n e^{-2\pi ilk/n} c_k\ .
\ee
This defines a twist around $w=u$ 
\be\label{replicabcOrb}
{\tilde b}_k(e^{2\pi i} (w-u)) = e^{-2\pi ik/n} {\tilde b}_k(w-u)\ ,\qquad
{\tilde c}_k(e^{2\pi i} (w-u)) = e^{2\pi ik/n} {\tilde c}_k(w-u)\ ,
\ee
and likewise an anti-twist around $w=v$.\
These are recognized as standard orbifold boundary conditions for conical 
singularities located at $u, v$. The conformal transformation argument 
for the stress tensor then leads to twist operators in the replica 
$n$-space with dimensions ${c\over 24} (1-{1\over n^2})$ with $c=-2$, 
as in our discussion in sec.~3: thus entanglement entropy in this CFT 
can be obtained as discussed there. Since these are Grassman 
variables, the treatment of the path integral is similar to that 
for fermions (discussed \eg\ in \cite{Casini:2005rm,Casini:2009sr}).

To understand the replica boundary conditions in some more detail, 
we consider a $\BZ_N$ orbifold of the $bc$-theory with a twist by 
$e^{2\pi ik/N}$ around the origin defined by
\be\label{twistbc}
b_t(e^{2\pi i} z) = e^{-2\pi ik/N}  b_t(z)\ ,\qquad 
c_t(e^{2\pi i} z) = e^{2\pi ik/N}  c_t(z)\ ,\qquad\qquad
k=1,\ldots, N-1\ ,
\ee
and likewise for the anti-holomorphic sector. The mode expansions for 
the fields implementing these twist boundary conditions can be written as
\be
b_t(z) = \sum_{m\in\BZ} {b_{m+k/N}\over z^{m+1+k/N}}\ ,\qquad
c_t(z) = \sum_{m\in\BZ} {c_{m-k/N}\over z^{m-k/N}}\ .
\ee
The twist ground state $|0\rangle_{k/N}$ is annihilated by all 
operators with positive mode number, \ie\ 
$b_{m+k/N} |0\rangle_{k/N} = 0,\ c_{m-k/N} |0\rangle_{k/N} = 0,\ \ m>0$.
The contour argument using the short distance behaviour as $z\ra w$ 
away from the singularity leads to the anti-commutation relations
\be
\{ b_{m+k/N},\ c_{n-k/N} \} = \delta_{m+n,0}\ .
\ee
Note that there is no zero mode in this orbifold twisted sector 
($k\neq 0$), although the untwisted sector retains the $b,c$ zero 
modes. 
The correlation function can be found using the mode expansions 
and the $b_{m+k/N},c_{m-k/N}$-operators as
\be\label{2ptfntwistbc}
\langle b_t(z) c_t(w)\rangle_{k/N} =
\sum_{m,n} {\langle 0| b_{m+k/N}\ c_{n-k/N} |0 \rangle_{k/N}\over 
z^{m+1+k/N} w^{n-k/N}} = \sum_{m=0}^\infty {1\over z}\ {w^{k/N}\over z^{k/N}}\ 
\Big({w\over z}\Big)^m = \Big({z\over w}\Big)^{-k/N} {1\over z-w}\ .
\ee
%where the zero mode being absent in the twisted sector allows the 
%normalization $\langle 0| 0\rangle_{k/N} = 1$.\
The position dependence of this correlation function reflects the 
expected properties under going around the conical singularity as well 
as the short distance behaviour\ ${1\over z-w}$\ as $z\ra w$ which 
is not affected by the boundary conditions.\ 
Likewise
\be\label{2ptfntwistbdc}
\langle b_t(z) \del c_t(w)\rangle_{k/N} =
\sum_{m,n} {\langle b_{m+k/N}\ c_{n-k/N} \rangle\over z^{m+1+k/N} w^{n+1-k/N}} 
%= \sum_{m=1}^\infty {m+{k\over N}\over zw}\ {z^{k/N}\over w^{k/N}}\ 
%\Big({w\over z}\Big)^m \nonumber\\
= {1\over z}\ \Big({z\over w}\Big)^{1-k/N}\ {{k\over N}z + (1-{k\over N})w
\over (z-w)^2}\ .
\ee
This is similar in form to the complex ($c=2$) boson correlation 
function in \cite{Dixon:1986qv}, with a sign difference 
reflecting the short distance ghost structure. 

In terms of the twist field $\sigma_{k/N}(z)$ creating the twist 
boundary conditions at location $z$, with appropriate OPEs that can 
be written as in \eg\ \cite{Dixon:1986qv}, the twist ground state 
is $|0\rangle_{k/N}=\sigma_{k/N}(z) |0\rangle$. 
The boundary conditions (\ref{twistbc}) can then be defined in terms 
of the twist field $\sigma_{k/N}(z)$ as
\be\label{twistk/N}
b_t(e^{2\pi i} z) \sigma_{k/N}(0) = e^{-2\pi ik/N}  b_t(z) \sigma_{k/N}(0)\ ,
\qquad 
c_t(e^{2\pi i} z) \sigma_{k/N}(0) = e^{2\pi ik/N}  c_t(z) \sigma_{k/N}(0)\ ,
\ee
with $k=1,\ldots, N-1$.\ This is equivalent to the OPEs
\be
b_t(z) \sigma_{k/N}(0) \sim z^{-k/N} \tau_{k/N} ,\qquad 
\del c_t(z) \sigma_{k/N}(0) \sim z^{-1+k/N} \tau'_{k/N}\ ,
\ee
etc, with $\tau_{k/N}, \tau'_{k/N}$ being excited twist fields.\ 
Corresponding to the twist field $\sigma_{k/N}\equiv \sigma^+_{k/N}$ 
here is the anti-twist field $\sigma^-_{k/N} \equiv \sigma^+_{1-k/N}$ 
defined via similar OPEs (or boundary conditions (\ref{twistk/N})) but 
with ${k\over N}\ra 1-{k\over N}$~. Thus the $bc$ correlation function 
(\ref{2ptfntwistbc}) in this twist sector can be written as\
$\langle 0| \sigma^-_{k/N} b_t(z) c_t(w) \sigma^+_{k/N}|0\rangle$ 
with the twist and anti-twist operator inserted.\ 
The expectation value of the energy-momentum tensor in this twist 
ground state can be obtained by regularizing (point-splitting) as
\be
\langle T(z) \rangle_{k/N} = \displaystyle{\lim_{z\to w}} 
\Big(\langle -:b_t(z) \del c_t(w):\rangle + {1\over (z-w)^2}\Big)
\ee
Expanding (\ref{2ptfntwistbdc}) in $z-w$, the ${1\over (z-w)^2}$ 
singularity cancels (effectively normal ordering $T(z)$) and we obtain
\be\label{twistdim}
\langle T(z) \rangle_{k/N} = 
-{1\over 2} {k\over N} \Big( 1 - {k\over N} \Big) {1\over z^2}\ \equiv\ 
{h_{\sigma_{k/N}}\over z^2}\ .
\ee
This gives the dimension of the twist field $\sigma_{k/N}(z)$, which 
we note is negative (this has also appeared previously in \eg\ 
\cite{Saleur:1991hk,Kausch:1995py}): this is a reflection of the short 
distance ghost behaviour and the negative central charge (and associated 
negative norm states).\ Likewise we can calculate the $U(1)$ charge 
of the twist field $\sigma^+_{k/N}$ as
\be
\langle j(z)\rangle_{k/N} = \displaystyle{\lim_{z\to w}} 
\Big(\langle -:b^t(z) c^t(w):\rangle + {1\over z-w}\Big) 
= {k/N\over z}\ ,
\ee
so that the $U(1)$ charge of $\sigma^+_{k/N}$ is ${k\over N}$\ 
(see also \cite{Saleur:1991hk,Kausch:1995py}). A similar calculation 
shows that the dimension of the anti-twist field $\sigma^-_{k/N}\equiv 
\sigma^+_{1-k/N}$ is $-{1\over 2} {k\over N} (1-{k\over N})$ while its
$U(1)$ charge is $1-{k\over N}$~. Correlation functions of twist field 
operators also require the total $U(1)$ charge to cancel the background 
charge when calculated in the untwisted $SL(2)$ vacuum for being 
nonvanishing, \ie\ 
\be\label{twistcorrfnQcharge}
\langle \sigma^+_{\lambda_1} \sigma^+_{\lambda_2}\ldots \rangle 
\equiv \langle 0| \sigma^+_{\lambda_1} \sigma^-_{1-\lambda_2}\ldots |0\rangle
\neq 0 \qquad \Rightarrow\qquad \sum_i\lambda_i = 1\ ,
\ee
reflecting the normalization (\ref{qnorm}). This is expected since for 
field insertions far from the region containing twist operator 
insertions, the OPEs resemble those in the untwisted theory so that 
correlation functions are nonvanishing only if the background charge is 
cancelled. In the bosonized formulation, we have\ $j(z) = i\del\phi$ and 
$b(z) = e^{-\phi} ,\ c(z) = e^{\phi} ,$ in the untwisted $c=-2$ theory. 
In the sector twisted by $\lambda={k\over N}$ the twist fields are\ 
$\sigma_\lambda = e^{i\lambda \phi} = \sigma^+_\lambda$.\ From this point 
of view it is clear in particular that a nonvanishing 2-point function 
is of the form\ 
$\langle 0| \sigma^+_{\lambda} \sigma^-_{\lambda} |0\rangle = 
\equiv \langle 0| \sigma^+_{\lambda} \sigma^+_{1-\lambda} |0\rangle$,\
which automatically contains an unpaired $c$-field $e^\phi$ cancelling
the background charge $Q=-1$.

Returning to the $n$-sheeted replica space defined by the boundary 
conditions (\ref{replicabc}), (\ref{replicabcdiag}), (\ref{replicabcOrb}), 
we note that the singularity at $w=u$ twists the $b$-field while that 
at $w=v$ anti-twists the $b$-field, and likewise for the $c$-field.
In terms of the diagonalized fields ${\tilde b}, {\tilde c}$, in 
(\ref{replicabcdiag}), the different sheets are decoupled: thus we can 
write the replica partition function as
\be
tr \rho_A^n\ =\ \prod_{k=1}^{n-1} \langle \sigma^-_{k/N}(v) 
\sigma^+_{k/N}(u) \rangle_0\ =\ (v-u)^{-4\sum_{k=1}^{n-1}\ h_{\sigma_{k/N}}}\ 
=\ (v-u)^{{1\over 3} (n-1/n)}\ ,
\ee
where the twist operator 2-point function is\ 
$\langle 0| \sigma^-_{k/N}(v) \sigma^+_{k/N}(u) |0 \rangle$ 
\ on sheet$-k$ 
for the $\lambda=1$ $bc$-theory we are focussing on with $c=-2$ in the 
$SL(2)$ vacuum (equivalently the ghost ground state in this case). 
Since the neighbourhood of each singularity does not contain zero modes,
one might be concerned about zero modes in the $n\ra 1$ limit: the 
normalization (\ref{twistcorrfnQcharge}) w.r.t. the background charge 
for twist field correlation functions (and more generally (\ref{qnorm})) 
shows that the $n\ra 1$ limit is smooth with regard to the correlation 
function being nonvanishing. This gives
\be
S_A = -\displaystyle{\lim_{n\to 1}} \ \del_n\ tr \rho_A^n\ =\ 
- {2\over 3} \log {l\over \epsilon}\ ,
\ee
where $l\equiv v-u$ and $\epsilon$ is the ultraviolet cutoff. This 
is of the standard form ${c\over 3} \log {l\over \epsilon}$ with 
$c=-2$ and is thus negative.

%For more general $bc$-ghost systems, the twist 2-point function in 
%the $SL(2)$ vacuum is\ 
%$\langle \sigma^-_{k/N}(v) \sigma^+_{k/N}(u) \rangle_0 = 
%\langle -Q|\ \sigma^-_{k/N}(v) \sigma^+_{k/N}(u)\ |0\rangle$\ with the 
%normalization (\ref{qnorm}) appropriate for the $SL(2)$ vacuum of the 
%ghost theory with background charge $Q$ which must be cancelled.

It is noteworthy that negative central charge in the present context 
leads to twist operators with negative conformal dimension $h_{\sigma_{k/N}}$ 
as we have seen (\ref{twistdim}). This leads to a 2-point function of
the form 
$\langle \sigma^-_{k/N}(v) \sigma^+_{k/N}(u) \rangle = |v-u|^{4(k/N)(1-k/N)}$.
More generally the replica formulation gives 
$\langle\Phi_n\Phi_{-n}\rangle \sim\ |v-u|^{|c|(1-1/n^2)/6}$.\ 
Thus there is no short distance divergence here as $|v-u|\ra 0$. 
However for large separations $|v-u|\ra\infty$, the 2-point function
diverges\footnote{This has also been noted in \cite{Caraglio:2008pk} 
who study entanglement entropy in certain $Q$-state Potts models with 
$c<0$, including numerical analysis. The $Q$-state Potts models, as 
statistical mechanical systems defined on an arbitrary lattice, exhibit 
a $Q$-valued permutation symmetry, the basic spin variable taking $Q$ 
values (the Ising model being $Q=2$), with 
spin-spin interactions. On a square lattice, with $Q$ varying between 
$0$ and $4$, these exhibit critical points described by CFTs with 
central charge\ $c=1-{6\over m(m+1)}$\ and\ 
$\sqrt{Q}=2\cos {\pi\over m+1}$~. For $Q<1$ (after appropriately 
defining the partition function), the CFT is non-unitary, with $c<0$.
In particular, for $Q=2-\sqrt{3}$, the CFT has central charge 
$c=-{11\over 14}$~. \cite{Caraglio:2008pk} note that negative 
entanglement entropy suggests that the mixed state corresponding to 
the reduced density matrix is apparently more ordered than the 
ground state which has vanishing entropy. As a check, they note 
that the correlation function of the branch-point twist fields 
(with conformal dimension proportional to $c<0$) grows with distance.}: 
this is an infrared divergence, and is reminiscent of long-distance
instabilities in the replica theory.  Note that in the limit $n\ra 1$,
the twist operators approach zero conformal dimension: this suggests
that the replica CFT instability is possibly ``marginal'' at worst.

Finally (although the details are quite different) our discussion 
appears consistent with \cite{Bianchini:2014uta} who argue that the
entanglement entropy has the form $c_{eff} \log {l\over\epsilon}$~. 
The effective central charge $c_{eff} = c-24\Delta$ is often positive, 
with $\Delta$ the (negative) dimension of the operator with lowest 
conformal dimension.  In the present $bc$-system with $c=-2$, the 
field of lowest conformal dimension is $c$ with dimension $\Delta=h_c=0$, 
so that $c_{eff} = c < 0$. It would be interesting to understand 
this better.

\subsection{Matter + ghost systems}

Let us now consider the $bc$-system (focussing on the $c=-2$ case)
alongwith some other CFT with central charge $c_m>1$: we have in mind
a free CFT of say $c_m$ free scalars. The total central charge of the
full system is $c_m+c_{bc}=c_m-2$. We assume the matter sector has a
well-defined $SL(2)$ invariant ground state $|0_m\rangle$ with
$L_0^m|0_m\rangle=0$ so that the full theory has a $SL(2)$ invariant
vacuum given by $|0_m\rangle\otimes |0_{bc}\rangle$, with\
$L_0^m+L_0^{bc} = 0$. Correlation functions in the ghost sector are 
defined with the normalizations (\ref{qnorm}) for $Q=-2$, while in 
the matter sector we assume $\langle 0_m| 0_m\rangle = 1$.\
Then from the previous arguments, the replica formulation for a single
interval amounts to independent calculations in the matter and ghost
sectors since the two sectors are free and decoupled from each
other. In particular,\ $tr\rho_A^n = \prod_k
\langle \sigma^-_{k/N}(v) \sigma^+_{k/N}(u) \rangle_m\ 
\langle \sigma^-_{k/N}(v) \sigma^+_{k/N}(u) \rangle_{bc}$\ with a 
product structure on the matter and ghost sectors.
This leads to
\be
S_A = {c_m - 2\over 3}\ \log {l\over\epsilon}\ ,
\ee
so that the entanglement entropy now effectively corresponds to that 
of a CFT with central charge $c_m-2$. Perhaps this is not surprising 
since the ghost system acts to cancel central charge and corresponding 
conformal anomalies. In the present context, this suggests that 
the ghost system ``cancels'' entanglement, \ie\ the ghosts effectively 
disentangle degrees of freedom of the original system. We mention 
that this again is formal: the physical interpretation of this 
entanglement cancellation is not clear at this point.

\section{A ghost\ $\chi{\bar\chi}$-CFT of anti-commuting scalars}

The first order $bc$-ghost system with $c=-2$ discussed above can be
thought of as a non-logarithmic sector of logarithmic conformal field 
theories studied in \cite{Gurarie:1993xq,Gurarie:1997dw,Gurarie:2013tma,
Kausch:1995py,Kausch:2000fu,Flohr:2001zs,Krohn:2002gh}. Consider a 
2-dim Euclidean CFT with action\ 
$S_E = \int d^2\sigma\ (\del_1\chi\del_1{\bar\chi} 
+ \del_2\chi\del_2{\bar\chi})$\
defined on the Euclidean plane $(\sigma^1,\sigma^2)$, consisting of 
two anticommuting complex massless scalars $\chi,\ {\bar\chi}$, 
regarded as Grassmann variables. In the present case, we are motivated 
by \cite{Anninos:2011ui} involving the 3-dim $Sp(N)$ Euclidean higher 
spin theory of anti-commuting scalars (see also 
\cite{LeClair:2006kb,LeClair:2007iy}): we consider the 2-dim ghost 
CFT here as a toy model, using standard CFT tools.
Structurally this $\chi{\bar\chi}$-CFT has some similarities with a 
complex commuting boson $\phi$ with action\ 
$S\sim \int d^2z (\del\phi{\bar\del}{\bar\phi} 
+ {\bar\del}\phi\del{\bar\phi})$, for instance in its stress tensor 
$T(z)\sim -\del\phi\del{\bar\phi}$,\ operator product expansions 
(OPEs) and mode expansions etc, but with some crucial differences due 
to the Grassmann nature. 

%A Lorentzian 
%continuation of this theory is obtained by taking $\sigma^1$ as 
%Euclidean time and rotating as $\sigma^1=it$. This gives\
%$S_E \ra -i\int dtd\sigma\ \big({\dot\chi} {\dot{\bar\chi}} 
%- \chi'{\bar\chi}' \big) \equiv -iS_{Mink}$.\ 
%This is analogous to a complex commuting boson continuing as\ 
%$S_E = \int d\tau d\sigma 
%(\del_\tau\phi \del_\tau{\bar\phi} + \phi' {\bar\phi}') \ra 
%-i\int dt d\sigma ({\dot\phi}{\dot{\bar\phi}} - \phi' {\bar\phi}') 
%= -iS_{Mink}$.

Defining\ $z=\sigma^1+i\sigma^2 ,\ {\bar z} =\sigma^1-i\sigma^2$, gives 
%complex coordinates\ %$\del\equiv \del_z = {1\over 2} (\del_1-i\del_2) , \
%{\bar\del}\equiv {\bar\del}_{{\bar z}} = {1\over 2} (\del_1+i\del_2)$ and
%\ $\del_1=\del+{\bar\del} ,\ \del_2=i(\del-{\bar\del})$. $d^2z=2d^2\sigma$.
the action and equations of motion 
\be\label{Schibarchi}
S = \int d^2z\ \big( \del \chi {\bar\del} {\bar\chi}\ 
+\ {\bar\del} \chi \del{\bar\chi} \big)\ ,\qquad\qquad 
{\bar\del} \del \chi = 0\ , \qquad {\bar\del} \del {\bar\chi} = 0\ ,
\ee
which then give the OPEs
\bea\label{chibarchiOPE}
&& \qquad\qquad\qquad\qquad 
\chi(z) {\bar\chi}(w)\ \sim\ -\log |z-w|^2\ ,\nonumber\\ 
{\ie} && \ \ 
\del_z\chi(z) \del_w{\bar\chi}(w)\ \sim\ -{1\over (z-w)^2}\ ,\qquad
\del_w{\bar\chi}(w) \del_z\chi(z)\ \sim\ {1\over (z-w)^2}\ .
\eea
This is similar to the commuting boson case, 
%The $\chi(z) {\bar\chi}(w)$ OPE differs by a minus sign from the 
%stemming from the minus sign that arises from the 
%Grassmann nature of $\chi, {\bar\chi}$\ (normalized as in the action). 
except for the relative minus sign in the second line due to the 
anti-commuting nature of $\chi, {\bar\chi}$.\ The presence of the two 
derivatives makes the $\del\chi\del{\bar\chi}$ OPE differ from that 
for $\del{\bar\chi}\del\chi$ by a minus sign, consistent with the 
short-distance singularity being ${-1\over (z-w)^2}$ rather than 
\eg\ ${1\over z-w}$ as in the $bc$-ghost CFT.
From the action\ $S\sim \int d^2\sigma \sqrt{g} g^{ij} 
\del_i\chi \del_j{\bar\chi}$, the energy-momentum tensor is
\be
T_{ij} \sim -{1\over\sqrt{g}} {\delta S\over \delta g^{ij}}\ \sim\ 
-\del_i\chi \del_j{\bar\chi} + {1\over 2} g_{ij} L
\quad \Rightarrow\qquad
T_{zz} \sim\ -\del\chi \del{\bar\chi}\ ,\qquad
T_{{\bar z}{\bar z}}\sim -{\bar\del}\chi {\bar\del}{\bar\chi}\ ,\quad
\ee
where we evaluate $T_{zz}={1\over 4} (T_{11}-T_{22}-2iT_{12})$ using 
complex coordinates, and so on (or via Noether's theorem).
It can be checked that $T_{z{\bar z}}=0$ which reflects conformal 
invariance of the theory.
The $TT$ OPE has the standard form with $c=-2$: we have
\be
T(z) T(w) =\ \ :\del\chi \del{\bar\chi} (z):\ :\del\chi \del{\bar\chi} (w):\ \ 
%\nonumber\\
%&=& -{1\over (z-w)^4}\ + \Big(-{1\over (z-w)^2}\Big) 
%:\del\chi(z) \del{\bar\chi} (w):\ + 
%\Big({1\over (z-w)^2}\Big) :\del{\bar\chi} (z) \del\chi(w): \nonumber\\
= -{1\over (z-w)^4} + {2\over (z-w)^2} T(w) 
+ {1\over (z-w)} \del T(w) . %\qquad\quad c=-2\ .
\ee
The $c=-2$ central term with the minus sign arises from moving 
${\bar\chi}$ through the other fields, since $\chi, {\bar\chi}$ 
anticommute. The OPE\ $T(z) \del\chi(0) \sim\ {1\over z^2} \del\chi(0) 
+ {1\over z} \del^2\chi(0)$\ of $T(z)$ with $\del\chi$\ (and likewise 
$\del{\bar\chi}$) is similar to that of a complex commuting $c=2$ 
scalar and show that $\del\chi$ and $\del{\bar\chi}$ have conformal 
dimension one.
From the equations of motion ${\bar\del}\del\chi=0,\ 
{\bar\del}\del{\bar\chi}=0$, we see that $\del\chi$ and $\del{\bar\chi}$ 
are holomorphic giving the mode expansions
\be\label{chiExp}
\del\chi = -i\sum_{m=-\infty}^\infty {\chi_m\over z^{m+1}}\ ,\qquad 
\del{\bar\chi} = -i \sum_{m=-\infty}^\infty {{\bar\chi}_m\over z^{m+1}}\ .
\ee
Inverting we have\ $\chi_m = \oint {dz\over 2\pi} z^m\del\chi ,\
{\bar\chi}_m = \oint {dz\over 2\pi} z^m\del{\bar\chi}$.
The OPEs (\ref{chibarchiOPE}) and likewise 
$\del_z\chi\del_w\chi\sim reg$ and $\del_z{\bar\chi} \del_w{\bar\chi}\sim reg$ 
can be used via the usual contour arguments, \eg\ 
$\{\chi_m, {\bar\chi}_n\} = \oint_{C_2} {dw\over 2\pi}\ 
{\rm Res}_{z\to w} z^m w^n \del_z\chi\del_w{\bar\chi}$.\ These give 
the anti-commutation relations for the oscillators,
\be\label{chibarchiOsc}
\{\chi_m, {\bar\chi}_n\} = m\delta_{m+n, 0}\ ,\qquad
\{\chi_m, \chi_n\} = 0\ ,\qquad \{{\bar\chi}_m, {\bar\chi}_n\} = 0\ .
\ee
In particular this means $\{\chi_0, {\bar\chi}_0\}=0$ due to the $m$ factor.\
From the mode expansions (\ref{chiExp}), we obtain (for the holomorphic 
parts)
\be
\chi(z)=\xi_0-i\chi_0\log z - i\sum_{m\neq 0} {\chi_m\over m z^m}\ ,
\qquad  {\bar\chi}(z)={\bar\xi}_0-i{\bar\chi}_0\log z 
- i\sum_{m\neq 0} {{\bar\chi}_m\over m z^m}\ ,
\ee
with $\xi_0,\ {\bar\xi}_0$ the constant parts or the zero modes of 
the $\chi,{\bar\chi}$-fields.
From this and the OPEs (\ref{chibarchiOPE}), we find the additional 
anti-commutation relations
\be
\{\xi_0, {\bar\chi}_0\} = i\ , \qquad \{{\bar\xi}_0, \chi_0\} = -i\ .
\ee
The presence of the zero modes $\xi_0, {\bar\xi}_0$, leads to 
similarities with the $bc$-ghost systems earlier, but with additional 
features here. 
In particular, there are logarithmic operators in this theory, \eg\
\be
{\tilde\BI} =\ :\chi{\bar\chi}:\ ,\qquad\qquad 
T(z) {\tilde\BI}(w) = {\BI\over (z-w)^2} + {1\over z-w} \del{\tilde\BI} 
+ \ldots\ ,
\ee
with $\BI$ the identity operator, and ${\tilde\BI}$ its logarithmic 
partner. Thus\ $[L_0, {\tilde\BI}] = \BI$, so that $L_0$ cannot be 
completely diagonalized, giving a 2-component ``Jordan cell'' with 
$L_0$ eigenvalue zero. These features give rise to logarithms in 
some correlation functions: the conditions on correlation functions 
following from conformal invariance lead to coupled differential equations 
whose solutions involve power-laws $z^\al$, as well as $z^\al\log z$\ 
(see \eg\ (\ref{corrnfnlog1}) later), as discussed in 
\cite{Gurarie:1993xq} and also \cite{Gurarie:1997dw,Gurarie:2013tma,
Kausch:1995py,Kausch:2000fu,Flohr:2001zs,Krohn:2002gh}.

Using the expansion $T(z)=\sum_m {L_m\over z^{m+2}}$ and the OPEs of 
$T(z)$ with $\del\chi,\ \del{\bar\chi}$, it can be seen that the 
subsector comprising excited states built over the ground state 
$|0\rangle$ by oscillators, \eg\ $\prod \chi_{-n}{\bar\chi}_{-m}|0\rangle$, 
have positive 
$L_0\sim \sum_n (-\chi_{-n}{\bar\chi}_n + {\bar\chi}_{-n}\chi_{n})$\ 
eigenvalue in this subspace (with $L_m\sim -\sum_n \chi_n {\bar\chi}_{m-n}$).
These are fermionic excitations since $\chi_k^2,\ {\bar\chi}_k^2=0$. 
It is useful to recall that for fermionic oscillator operators 
satisfying $\{ a, a^\dag \} = 1$, a Hamiltonian $H=a^\dag a$ gives 
$[H, a] = -a,\ [H, a^\dag] = a^\dag$ so that one can define a 
ground state as the lowest weight state with $a|vac\rangle = 0$. 
However for oscillators with $\{a, a^\dag\} = -1$, we have 
$[a^\dag a, a] = a,\ [a^\dag a, a^\dag] = -a^\dag$. However defining 
$H=-a^\dag a$ gives $[H, a] = -a,\ [H, a^\dag] = -a^\dag$ allowing 
for excited states built over the ground state regarded as a lowest 
weight state.\ In the present case, the subsector built over the vacuum 
$|0\rangle$ has a representation for $L_0$ which gives $|0\rangle$ 
as a lowest weight state, with\
$[L_0, \chi_{-n}]=n\chi_{-n} ,\ [L_0, {\bar\chi}_{-n}]=n{\bar\chi}_{-n} ,\
[L_0, \chi_{n}]=-n\chi_{n} ,\ [L_0, {\bar\chi}_{n}]=-n{\bar\chi}_{n}$,\ 
$(n>0)$, consistent with $\chi_{-n},\ {\bar\chi}_{-n},\ n>0$ being 
creation operators, and $\chi_n,\ {\bar\chi}_n,\ n>0$ being 
annihilation operators.

Defining the ``logarithmic'' state
\be
|\xi_0\rangle \equiv \xi_0{\bar\xi}_0 |0\rangle\ ,
\ee
we see that due to the Grassmann integration over the zero modes 
$\xi_0, {\bar\xi}_0$,\ the vacuum of this theory $|0\rangle$ satisfies\ 
\be\label{0xi_0norm}
\langle 0| 0\rangle = 0\ ,\qquad 
\langle \xi_0| 0\rangle = \langle 0| \xi_0{\bar\xi}_0 |0\rangle = 1\ ,
\ee
where inserting the zero modes serves to give a nonzero answer in the 
Grassmann integration. From the path integral point of view, we have\
$\langle 0| 0 \rangle = \int D\chi D{\bar\chi}\ e^{-S} = 0$\ whereas\
$\langle \xi_0| 0\rangle = \int D\chi D{\bar\chi}\ 
\chi {\bar\chi}\ e^{-S} = 1$.\
Inserting a single logarithmic operator ${\tilde\BI}$ serves to ensure 
nonvanishing results for correlation functions of derivative operators 
built out of $\del\chi, \del{\bar\chi}$\ (\ref{chiExp}), which do not 
contain the zero modes: this is a minimal way to restrict to a 
nonlogarithmic subsector of this theory. In particular, 
\be
\langle \xi_0|\ \del\chi(z) \del{\bar\chi}(w)\ |0\rangle \equiv 
\langle 0|\xi_0{\bar\xi}_0\ \del\chi(z) \del{\bar\chi}(w)\ |0\rangle 
= - {1\over (z-w)^2} \ .
\ee
Thus $|\xi\rangle = \xi_0{\bar\xi}_0 |0\rangle$ can be regarded as 
the out-state for the in-state being the vacuum $|0\rangle$,\ 
analogous to (\ref{qnorm}) for the $bc$-ghost system. With the norms 
of states defined in this manner, correlation functions of derivative 
operators (which do not contain the zero modes) have form similar to 
those of the commuting boson, with no additional logarithmic structure. 
We can use the mode expansions (\ref{chiExp}) to obtain the 2-point 
function as\
$\langle \xi_0| (-\sum_{n>0} n {w^{n-1}\over z^{n+1}}) |0\rangle 
= -\langle \xi_0|0\rangle {1\over z^2} \sum_{n=1}^\infty 
n {w^{n-1}\over z^{n-1}} = -{1\over (z-w)^2}$\ with the normalization 
(\ref{0xi_0norm}) for the ``inner product''.

Likewise we see negative norm states using (\ref{0xi_0norm}) and
defining hermiticity relations in the following way: by analogy with
the complex boson, it is consistent to take the hermitian conjugate
field to $\chi$ as\ $\chi^\dag = {\bar\chi}$ which gives\ 
$T(z)^\dag = T(z)$.
In the Euclidean theory, the hermitian conjugate can be defined as\ 
$\del\chi(z)^\dag\equiv \del{\bar\chi}({\bar z})$\ where 
$\phi(z)^\dag \equiv \phi^\dag({1\over {\bar z}}) {1\over {\bar z}^{2h}}$ 
with $h$ the conformal dimension of $\phi$.\ This gives\ 
$-i\sum_n {{\bar\chi}_n\over {\bar z}^{n+1}} = 
\sum_n i{\chi_n^\dag\over {\bar z}^{-n+1}} = 
\sum_n i {\chi_{-n}^\dag\over {\bar z}^{n+1}}$\ 
as usual for holomorphic field mode expansions, suggesting the 
hermitian conjugate operators\ 
$\chi_n^\dag=-{\bar\chi}_{-n} ,\ \chi_{-n}^\dag=-{\bar\chi}_n \ (n>0)$ 
and $\chi_0^\dag=-{\bar\chi}_0$.\ This gives $L_n^\dag = L_{-n}$ which 
is equivalent to $T(z)^\dag = T(z)$.\ 
This is consistent with the oscillator algebra (\ref{chibarchiOsc}),\
$\{\chi_m, {\bar\chi}_{-m}\} = m$.\ Then with the norms defined as 
(\ref{0xi_0norm}), we see that states of the form\
${\bar\chi}_{-m}|0\rangle,\ m>0$ have negative norm, since\ 
$\langle \xi_0| (-\chi_m) {\bar\chi}_{-m}|0\rangle = -m<0$. States of 
the form $\chi_{-m}|0\rangle$ have positive norm: more generally, 
states like $\prod \chi_{-m_i} {\bar\chi}_{-n_j}|0\rangle$ have 
negative norm for $\sum_i m_i < \sum_j n_j$\ (but positive $L_0$ 
eigenvalue).

\vspace{1mm}

\noindent {\bf Entanglement entropy:}\ \ We can formulate entanglement 
entropy through the replica as in the $c=-2$ $bc$-ghost system (sec.~3.2): 
a possible concern is the presence of logarithmic operators, as 
discussed in \cite{Gurarie:1993xq,Gurarie:1997dw,Gurarie:2013tma,
Kausch:1995py,Kausch:2000fu,Flohr:2001zs,Krohn:2002gh}, which can 
lead to further modifications to the logarithmic behaviour in the 
entanglement entropy. 
Before discussing this, we note that the $n$-sheeted replica theory is 
defined by boundary conditions 
\bea\label{replicachibarchi}
\chi_k(e^{2\pi i} (w-u)) = \chi_{k+1}(w-u)\ , &&
{\bar\chi}_k(e^{2\pi i} (w-u)) = {\bar\chi}_{k+1}(w-u)\ ,\qquad  
k=1,\ldots,n-1 , \ \ \nonumber\\
\chi_k(e^{2\pi i} (w-v)) = \chi_{k-1}(w-v)\ , && 
{\bar\chi}_k(e^{2\pi i} (w-v)) = {\bar\chi}_{k-1}(w-v)\ ,
\eea
with the $n$-th sheet connecting back to the first. Encircling both 
$w=u$ and $w=v$, gives trivial boundary conditions, \ie\ far from the 
interval the replica space is essentially a single sheet.
The fields and boundary conditions above can be diagonalized by defining\
%\be\label{replicachibarchidiag}
${\tilde \chi}_k = {1\over n} \sum_{l=1}^n e^{-2\pi ilk/n} \chi_k ,\  %\qquad
{\tilde {\bar\chi}}_k = {1\over n} \sum_{l=1}^n e^{2\pi ilk/n} {\bar\chi}_k$.
This defines a twist  %\be\label{replicachibarchiOrb}
${\tilde \chi}_k(e^{2\pi i} (w-u)) = e^{-2\pi ik/n} {\tilde \chi}_k(w-u) ,\
{\tilde {\bar\chi}}_k(e^{2\pi i} (w-u)) = 
e^{2\pi ik/n} {\tilde {\bar\chi}}_k(w-u)$,\ around $w=u$, 
and likewise an anti-twist around $w=v$.\
These are standard orbifold boundary conditions for conical 
singularities located at $u, v$. 
To illustrate this, consider a $\BZ_N$ orbifold of the 
$\chi{\bar\chi}$-theory with a twist by $e^{2\pi ik/N}$ around the origin 
\be\label{twistchibarchi}
\chi(e^{2\pi i} z) = e^{2\pi ik/N}  \chi(z)\ ,\qquad 
{\bar\chi}(e^{2\pi i} z) = e^{-2\pi ik/N}  {\bar\chi}(z)\ ,\qquad\qquad
k= 1,\ldots, N-1\ ,
\ee
and likewise for the anti-holomorphic sector. The mode expansions for 
the fields implementing these twist boundary conditions and the corresponding 
anti-commutation relations are
\be
\del\chi = \sum_m {\chi_{m-k/N}\over z^{m+1-k/N}}\ ,\ \ \
\del{\bar\chi} = \sum_m {{\bar\chi}_{m+k/N}\over z^{m+1+k/N}}\ ,\qquad
\{ \chi_{m-k/N},\ {\bar\chi}_{m+k/N} \} 
= \big( m - {k\over N} \big) \delta_{m+n,0}~.\ 
\ee
The twist ground state $|0\rangle_{k/N}$ is annihilated by all operators 
with positive mode number, \ie\ $\chi_{m-k/N} |0\rangle_{k/N} = 0$, 
and ${\bar\chi}_{m+k/N} |0\rangle_{k/N} = 0,\ m\geq 0$, and the 
anti-commutation relations 
are obtained by the contour argument using the short distance behaviour 
as $z\ra w$ away from the singularity. The twist ground state can be 
thought of as $|0\rangle_{k/N}=\sigma_{k/N}(z)|0\rangle$, with 
$\sigma_{k/N}(z)$ the twist field creating the twist boundary conditions 
at location $z$. The correlation function can be found using these 
mode expansions and the $\chi_{m-k/N}, {\bar\chi}_{m+k/N}$-operators as
\be\label{2ptfntwist}
\langle \del\chi(z) \del{\bar\chi}(w)\rangle_{k/N} = 
\sum_{m,n} {\langle \chi_{m-k/N}\ {\bar\chi}_{n+k/N} \rangle\over 
z^{m+1-k/N} w^{n+1+k/N}} 
%= \sum_{m=1}^\infty {m-{k\over N}\over zw}\ {z^{k/N}\over w^{k/N}}\ 
%\Big({w\over z}\Big)^m \nonumber\\
= {1\over z}\ \Big({z\over w}\Big)^{k/N} {(1-{k\over N})z + {k\over N}w
\over (z-w)^2}\ .
\ee
This is similar in form to the ($c=2$) complex boson correlation 
function in \cite{Dixon:1986qv}, except for a minus sign as expected 
for these anti-commuting $c=-2$ scalars: it is also similar in form 
to (\ref{2ptfntwistbdc}) in the $bc$-ghost system discussed previously. 
The expectation value of the energy-momentum tensor in this twist 
ground state can be obtained by regularizing as\
$\langle T(z) \rangle_{k/N} = \lim_{z\to w} 
(\langle -\del\chi(z) \del {\bar\chi}(w)\rangle + {1\over (z-w)^2})$, 
and we obtain
\be
\langle T(z) \rangle_{k/N} = 
-{1\over 2} {k\over N} \Big( 1 - {k\over N} \Big) {1\over z^2}\ \equiv\ 
{h_{\sigma_{k/N}}\over z^2}\ .
\ee
This gives the dimension of the twist field $\sigma_{k/N}(z)$. Not 
surprisingly this is the same as that for the complex boson, except 
for the minus sign.

As mentioned earlier, the presence of the logarithmic operators in 
these logarithmic conformal field theories (equivalently the fact 
that $L_0$ is not completely diagonalizable) leads to logarithms in 
correlation functions, rather than simple power laws alone, as 
discussed in \cite{Gurarie:1993xq,Gurarie:1997dw,Gurarie:2013tma,
Kausch:1995py,Kausch:2000fu,Flohr:2001zs,Krohn:2002gh}. In particular 
inserting multiple logarithmic operators leads to logarithmic behaviour 
in correlation functions, with \eg\ 
$\langle {\tilde \BI}(z) {\tilde \BI}(w) \rangle \sim -\log(z-w)$.\ 
The OPEs of twist field operators in these theories also contain 
logarithms, \eg\ 
\be\label{corrnfnlog1}
\mu(z) \mu(0) \sim z^{1/4} \big(\BI \log z\ +\ {\tilde \BI}\big) 
\ee
with a $Z_2$ twist, where the twist fields $\mu$ have conformal 
dimension $-{1\over 8}$~. More generally, for orbifolds with twists 
of the form (\ref{twistchibarchi}), we have \cite{Kausch:2000fu}
\be\label{corrnfnlog2}
\sigma^-_{k/N}(z) \sigma^+_{k/N}(w)\ \sim\ z^{-2h_{\sigma_{k/N}}} 
\big(\BI \log (z-w)\ +\ {\tilde \BI}\big)\ .
\ee
A 2-point correlation function in the vacuum $|0\rangle$ takes the form
\cite{Kausch:2000fu}
\be
\langle 0|\ \sigma^-_{k/N}(z,{\bar z}) \sigma^+_{k/N}(w,{\bar w})\ |0\rangle\ 
\sim\ |z-w|^{-4h_{\sigma_{k/N}}} = |z-w|^{2(k/N)(1-k/N)}\ ,
\ee
where only the term with the single logarithmic operator ${\tilde \BI}$ 
in the OPE contributes: this is in accord with the norm (\ref{0xi_0norm}). 
Thus the 2-point function has power law behaviour, with no logarithm.
The correlation function\ 
$\langle \sigma^-_{k/N}(z_1) \sigma^+_{k/N}(z_2) {\tilde \BI}(z_3) \rangle$\ 
exhibits logarithmic behaviour, with a form\ 
$|z_{12}|^{2(k/N)(1-k/N)} (\al + \beta \log |{z_{13}z_{23}\over z_{12}}|^2)$ 
with appropriate coefficients $\al, \beta$.\
More generally, higher point correlation functions of twist field 
operators (arising in \eg\ entanglement entropy for multiple disjoint 
intervals, or mutual information) are expected to exhibit the 
logarithms characteristic of these log-CFTs.

In terms of the diagonalized fields ${\tilde\chi}$, we see that the
different sheets are decoupled and thus we can write the replica
partition function as a product of 2-point functions of twist operators,
\be
tr \rho_A^n\ =\ \prod_{k=0}^{n-1} \langle \sigma^-_{k/N}(v) 
\sigma^+_{k/N}(u) \rangle\ =\ (v-u)^{-4\sum_{k=0}^{n-1}\ h_{\sigma_{k/N}}}\ 
=\ (v-u)^{{1\over 3} (n-1/n)}\ .
\ee
This gives
\be
S_A = -\displaystyle{\lim_{n\to 1}} \ \del_n\ tr \rho_A^n\ =\ 
- {2\over 3} \log {l\over \epsilon}\ ,
\ee
where $l\equiv v-u$ and $\epsilon$ is the ultraviolet cutoff. This 
is of the standard form ${c\over 3} \log {l\over \epsilon}$ with 
$c=-2$ and is negative. This is similar to the $bc$-ghost system which 
is in a sense a non-logarithmic subsector of the logarithmic CFT here.

%\vspace{2mm}

\section{A toy model of ``ghost-spins''}

To abstract away from the specific technical issues of the $bc$-ghost 
system, let us consider a very simple toy model of ``ghost-spins'' 
below, which mimics some of the key features. Firstly, for ordinary 
spin variables with a 2-state Hilbert space consisting of\ 
$\{ \uparrow,\ \downarrow \}$, we take the usual positive definite norms 
in the Hilbert space\ $\langle \uparrow | \uparrow\rangle = 
\langle \downarrow | \downarrow\rangle = 1$\ and\ 
$\langle \uparrow | \downarrow\rangle = 
\langle \downarrow | \uparrow\rangle = 0$.\ Then a generic state is\
$|\psi\rangle = c_1 | \uparrow\rangle + c_2 | \downarrow\rangle$, with 
adjoint\ $\langle\psi| = c_1^* \langle\uparrow| + c_2^* \langle\downarrow|$\ 
and norm\ $\langle \psi| \psi\rangle = |c_1|^2 + |c_2|^2$, which is positive 
definite. Thus we can normalize states as\ $\langle \psi| \psi\rangle = 1$
and pick a representative ray with unit norm (equivalent to calculating 
expectation values of operators as\ $\langle O \rangle 
= {\langle \psi| O |\psi\rangle \over \langle \psi| \psi\rangle} = 
\langle \psi| O |\psi\rangle$).\ 
The reduced density matrix obtained by tracing out the second spin is\
\be\label{redDM}
\rho_A = tr_B |\psi\rangle \langle \psi| = \sum_i \langle i_B|\psi\rangle 
\langle \psi| i_B\rangle = \langle \uparrow_B|\psi\rangle 
\langle \psi| \uparrow_B\rangle 
+ \langle \downarrow_B|\psi\rangle \langle \psi| \downarrow_B\rangle\ .
\ee
The familiar discussions in 2-spin systems of entanglement entropy 
via the reduced density matrix are recovered as follows.
States of the system such as\ \ $|\psi\rangle = c_1 |\uparrow \uparrow\rangle 
+ c_2 |\downarrow \downarrow\rangle$\ can be normalized as\
$\langle \psi| \psi\rangle = 1 = |c_1|^2 + |c_2|^2$ which is 
positive definite, and ensure that $|c_1|, |c_2| \leq 1$.\ With 
these norms, the reduced density matrix (\ref{redDM}) becomes
$\rho_A = |c_1|^2 | \uparrow\rangle \langle \uparrow |\ +\ 
|c_2|^2 | \downarrow\rangle \langle \downarrow |$.\ Note that 
the reduced density matrix is automatically normalized as\ 
$tr \rho_A = 1$ once the state $|\psi\rangle$ is normalized. Thus 
the entanglement entropy given as the von Neumann entropy of $\rho_A$ 
is\ $S_A = -tr \rho_A \log \rho_A = -\sum_i \rho_A(i) \log \rho_A(i)$ 
is positive definite since each eigenvalue $\rho_A(i) < 1$ makes 
the $-\log\rho_A(i) > 0$.

We define a single ``ghost-spin'' by a similar 2-state Hilbert space 
consisting of\ $\{ \uparrow,\ \downarrow \}$, but defining the ``norms'' 
as
\be\label{ghost-norms}
\langle \uparrow | \uparrow\rangle = 
\langle \downarrow | \downarrow\rangle = 0\ ,\qquad\quad
\langle \uparrow | \downarrow\rangle = 
\langle \downarrow | \uparrow\rangle = 1\ .
\ee
This is akin to the normalizations (\ref{qnorm}) in the $bc$-ghost 
system (see also \cite{polchinskiTextBk}, Appendix, vol.~1 where 
this inner product appears). Now a generic state 
\be
|\psi\rangle = c_1 |\uparrow\rangle + c_2 |\downarrow\rangle \qquad 
\Rightarrow\qquad \langle \psi| \psi \rangle = c_1 c_2^* + c_2 c_1^*\ ,
\ee
which is not positive definite: for instance the state\ 
$|\uparrow\rangle - |\downarrow\rangle$ has norm $-2$.\ It is then 
convenient to change basis to
\be\label{ghostbasis+-}
|\pm\rangle \equiv {1\over \sqrt{2}} \big(|\uparrow\rangle\ \pm\
|\downarrow\rangle \big)\ ,
\qquad
\langle +| + \rangle = 1\ ,\quad \langle -| - \rangle = -1\ ,
\quad \langle +| - \rangle = \langle -| + \rangle = 0\ .
\ee
A generic state with nonzero norm can be normalized to norm $+1$ or $-1$. 
Then a negative norm state can be written as
\be
|\psi\rangle = {1\over \sqrt{|c_2|^2 - |c_1|^2}} 
\big( c_1 |+\rangle + c_2 |-\rangle \big)\ ,\qquad 
\langle \psi| \psi\rangle = -1\ \qquad  \big(|c_2|^2 > |c_1|^2\big)\ .
\ee
For every state (or ray)\ $|\psi\rangle$ with norm $-1$, there is a 
corresponding state (or ray)\ $|\psi^\perp\rangle$ with norm $+1$ which 
is orthogonal to $|\psi\rangle$, 
\be
|\psi^\perp\rangle =  {1\over \sqrt{|c_2|^2 - |c_1|^2}} 
\big( c_2^* |+\rangle + c_1^* |-\rangle \big)\ ,\qquad 
\langle \psi^\perp| \psi^\perp\rangle = 1\ ,\qquad 
\langle \psi^\perp| \psi\rangle = 0\ .
\ee
There are also zero norm states given by
\be
|\psi\rangle = c_1 |+\rangle + c_2 |-\rangle\ ,\qquad 
\langle \psi| \psi\rangle = 0\ ,\qquad |c_2|^2 = |c_1|^2\ ,
\ee
which do not admit any canonical normalization, simple examples 
being $|\uparrow\rangle,\ |\downarrow\rangle$.

Now considering the two ghost-spin system, basis states are
\be\label{basisstates+-}
|s_A s_B\rangle\ \equiv\ 
|\uparrow \uparrow\rangle , \ |\uparrow \downarrow\rangle ,\ 
|\downarrow \uparrow\rangle , \ |\downarrow \downarrow\rangle\ \ \ \ 
\equiv\ \ \ | + + \rangle ,\ | + - \rangle ,\ | - + \rangle ,\ 
| - - \rangle\ .
\ee
The $|\pm\pm\rangle$ basis is more transparent for our purposes. 
Since the inner product or metric on this space of states is not 
positive definite, we need to be careful in defining the various 
contractions arising in the norms and partial traces. We define 
the states and norms as
\be
|\psi\rangle = \sum \psi_{ij} |ij\rangle\ ,\qquad 
\langle\psi|\psi\rangle\ \equiv\ g^{ik} g^{jl} \psi_{ij} \psi_{kl}^*\ =\ 
g^{ii} g^{jj} |\psi_{ij}|^2 \ ,
\ee
where repeated indices as usual are summed over: the last expression 
pertains to the $|\pm\rangle$ basis where the metric is diagonal, 
with\ $g^{++}=1,\ g^{--}=-1$.\ Thus a simple example of a positive 
norm state is\
$|\psi\rangle = \psi_{++} | + + \rangle\ + \psi_{--} | - - \rangle$\ 
with\ $\langle\psi| \psi\rangle = |\psi_{++}|^2 + |\psi_{--}|^2>0$ ,\
while\ $|\psi\rangle =  \psi_{+-} | + - \rangle\ + \psi_{-+} | - + \rangle$\ 
is a negative norm state with\ 
$\langle\psi| \psi\rangle = -(|\psi_{+-}|^2 + |\psi_{+-}|^2)<0$.\ 
Then 
\bea\label{psiijNorm1}
&& |\psi\rangle = {1\over \sqrt{c^2}} 
\big( c_1 | + + \rangle\ + c_2 | + - \rangle\ 
+ c_3 | - + \rangle\ + c_4 | - - \rangle  \big) \nonumber\\ 
&& \qquad\Rightarrow\qquad 
\langle\psi| \psi\rangle = {1\over c^2} \left( |c_1|^2 + |c_4|^2 - 
|c_2|^2 - |c_3|^2 \right) ,
\eea
is a generic state, so that normalized positive/negative norm states have 
\be\label{psiijNorm2}
c^2 = \pm  \left( |c_1|^2 + |c_4|^2 - |c_2|^2 - |c_3|^2 \right) > 0 
\qquad\Rightarrow\qquad \langle\psi| \psi\rangle = \pm 1\ ,
\ee
with norm $\pm 1$. This translates to corresponding conditions on the 
coefficients $c_i$.

With the density matrix $\rho=|\psi\rangle\langle\psi| 
= \sum \psi_{ij} \psi_{kl}^* | ij \rangle\langle kl|$, the reduced 
density matrix obtained by a partial trace over one spin can again be 
defined via a partial contraction as
\be
\rho_A = tr_B \rho \equiv (\rho_A)_{ik} |i\rangle \langle k|\ ,\qquad\ \  
(\rho_A)_{ik}\ =\ g^{jl} \psi_{ij} \psi_{kl}^*\ =\ 
g^{jj}  \psi_{ij} \psi_{kj}^*\ .
\ee
This gives
\bea
(\rho_A)_{++} =\ |\psi_{++}|^2 - |\psi_{+-}|^2\ , &\quad&
(\rho_A)_{+-} =\ \psi_{++} \psi_{-+}^* - \psi_{+-} \psi_{--}^*\ , \nonumber\\
(\rho_A)_{-+} =\ \psi_{-+} \psi_{++}^* - \psi_{--} \psi_{+-}^*\ , &\quad& 
(\rho_A)_{--} =\ |\psi_{-+}|^2 - |\psi_{--}|^2\ , 
\eea
For the state (\ref{psiijNorm1}), (\ref{psiijNorm2}), this is
\bea\label{rhoAc1234}
\rho_A &=& {1\over c^2} \Big[ \big(|c_1|^2 - |c_2|^2\big) |+ \rangle\langle +|\ 
+\ \big(c_1c_3^* - c_2 c_4^*\big) |+ \rangle\langle -|\ \nonumber\\
&& \qquad +\ \big(c_3c_1^* - c_4 c_2^*\big) |- \rangle\langle +|\ 
+\ \big(|c_3|^2 - |c_4|^2\big) |- \rangle\langle -| \Big]\ .
\eea
Then $tr \rho_A = g^{ik} (\rho_A)_{ik} = (\rho_A)_{++} - (\rho_A)_{--}$. 
Thus the reduced density matrix is normalized to have\ 
$tr \rho_A = tr \rho = \pm 1$ depending on whether the state 
(\ref{psiijNorm1}), (\ref{psiijNorm2}), is positive or negative norm.
Also, $\rho_A$ has some eigenvalues negative.

The entanglement entropy calculated as the von Neumann entropy of 
the reduced density matrix is
\be\label{EErhoA}
S_A = -g^{ij} (\rho_A \log \rho_A)_{ij}\ =\ - g^{++} (\rho_A \log \rho_A)_{++}
- g^{--} (\rho_A \log \rho_A)_{--}
\ee
where the last expression pertains to the $|\pm\rangle$ basis with 
$g^{\pm\pm}=\pm 1$. This requires defining $\log\rho_A$ as an 
operator\footnote{I thank D. Jatkar, A. Maharana and A. Sen for 
useful discussions here.}: we define this as 
\be
(\log\rho_A)_{ik} = (\log(1+\rho_A-1))_{ik} = 1_{ik} + (\rho_A-1)_{ik} 
+ (\rho_A-1)_{ij}g^{jl}(\rho_A-1)_{lk} + \ldots
\ee
or equivalently as the solution to\ $(\rho_A)_{ik} = (e^{\log\rho_A})_{ik} = 
1_{ik} + (\log\rho_A)_{ik} + (\log\rho_A)_{ij}g^{jl}(\log\rho_A)_{lk} + \ldots$.
\ The signs in the contractions in $\log\rho_A$ are perhaps more 
easily dealt with if we use the mixed-index reduced density matrix 
$(\rho_A)^i{_k}$.

To illustrate this, let us for simplicity consider a simple family of 
states where the reduced density matrix is diagonal, by restricting to 
$c_3^* = {c_2 c_4^*\over c_1}$. In this case, $\log\rho_A$ is also 
diagonal and can be calculated easily. From (\ref{rhoAc1234}) for the 
state (\ref{psiijNorm1}), (\ref{psiijNorm2}), this gives
\bea
c_3^* = {c_2 c_4^*\over c_1}\quad && \Rightarrow\qquad 
c^2 = \pm \left( |c_1|^2 - |c_2|^2 \right) \left( 1 + {|c_4|^2\over |c_1|^2} 
\right) > 0\ , \nonumber\\
&& \quad
\rho_A = \pm \left[ {|c_1|^2\over |c_1|^2 + |c_4|^2} |+ \rangle\langle +|\
-\ {|c_4|^2\over |c_1|^2 + |c_4|^2} |- \rangle\langle -| \right]\ ,
\eea
where the $\pm$ refer to positive and negative norm states respectively.
The location of the negative eigenvalue is different for positive and 
negative norm states, leading to different results for the von Neumann 
entropy. For negative norm states, we see that $c^2>0$ implies
$|c_2|^2>|c_1|^2$ and $|c_3|^2>|c_4|^2$, so that 
$(\rho_A)_{++} < 0,\ (\rho_A)_{--} > 0$.\
Then the mixed-index reduced density matrix components\ 
$(\rho_A)^i_k = g^{ij} (\rho_A)_{jk}$\ are
\be
(\rho_A)^+_+ = \pm x\ ,\qquad (\rho_A)^-_- = \pm (1-x)\ ,\qquad\quad 
x = {|c_1|^2\over |c_1|^2 + |c_4|^2}\ ,\qquad 0 < x < 1\ .
\ee
Thus we see that\ $tr\rho_A = (\rho_A)^+_+ + (\rho_A)^-_- = \pm 1$ 
manifestly. Now we obtain
\be
(\log\rho_A)^+_+ = \log (\pm x)\ ,\qquad 
(\log\rho_A)^-_- = \log (\pm (1-x))\ ,
\ee
the $\pm$ referring again to positive/negative norm states respectively. 
Thus the entanglement entropy (\ref{EErhoA}) becomes
\be
S_A = - (\rho_A)^+_+ (\log\rho_A)^+_+ - (\rho_A)^-_- (\log\rho_A)^-_-\ 
\ee
For positive norm states, we obtain
\be
\langle\psi| \psi\rangle > 0:\qquad S_A = - x\log x - (1-x) \log (1-x) > 0\ ,
\ee
which is manifestly positive since $x<1$, just as for the familiar 
entanglement entropy in an ordinary 2-spin system.\ For negative norm 
states however, we have
\be
\langle\psi| \psi\rangle < 0:\qquad S_A =  x\log (-x) + (1-x) \log (-(1-x)) 
= x\log x + (1-x) \log (1-x) + i\pi\ .\ \ \
\ee
We note that the imaginary part (using $\log (-1)=i\pi$) is independent 
of $x$, \ie\ the same for all such negative norm states.
The real part of entanglement entropy is negative since $x<1$ and 
the logarithms are negative: apart from the minus sign, it is the same 
as $S_A$ for the positive norm states. This real part is minimized 
when $x={1\over 2}$\ (this value corresponds to maximal entanglement 
for positive norm states): this ``minimal'' entanglement is 
$S_A = -\log 2 + i\pi$.

The above discussion can also be phrased in terms of the $|\uparrow\rangle, 
|\downarrow\rangle$ basis although we have found it convenient to use 
the $|\pm\rangle$ basis. It is worth noting that while (\ref{ghost-norms}) 
mimics the ghost norms (\ref{qnorm}), there is no obvious analog of 
the background charge here: in particular tracing over spin$_A$ instead 
of spin$_B$ is equivalent, so that entanglement entropy for the 
subsystem is the same as that for the complement.

It would appear that this discussion can be generalized to an
arbitrary lattice $L$ containing ghost-spins $|\uparrow_i\rangle,
|\downarrow_i\rangle$ at each lattice site $i\in L$. Using the
$|\pm_i\rangle$ basis, states with negative norm can be constructed as
above, and a subsystem can presumably be defined as a connected
spatial subregion in the lattice. It would be interesting to explore
entanglement entropy in these ghost-spin systems more completely, as
well as coupling to ordinary spin systems.

\section{Discussion}

We have studied entanglement entropy in some Euclidean non-unitary
CFTs with a view to gaining more insight into the areas of the complex
extremal surfaces on a constant Euclidean time slice in de Sitter
space (Poincare slicing) studied in
\cite{Narayan:2015vda,Narayan:2015oka}. In particular for $dS_4$, the
areas are negative reflecting the negative central charge
$-{R_{dS}^2\over G_4}$. With a view to finding toy models exhibiting
negative entanglement entropy, we have studied 2-dim ghost CFTs with
negative central charge, revisiting the replica formulation of
entanglement entropy for a single interval. In particular in the
$bc$-ghost system with $c=-2$, the $SL(2)$ vacuum coincides with the
ghost ground state enabling the application of the replica
formulation, which yields negative entanglement entropy. This also
involves appropriate inner products for states necessary for
nonvanishing correlation functions. Similar discussions apply for a
CFT of anti-commuting scalars which is a logarithmic CFT. We have also
discussed a toy model of two ``ghost-spins'' with non-positive inner
products mimicking some of these features, where the reduced density
matrix gives von Neumann entropy for negative norm states with a
negative real part as well as a constant imaginary part. 

Our analysis here illustrates that negative EE can formally be
obtained from appropriate generalizations of the standard CFT replica
technique in toy 2-dim ghost CFTs with negative central charge: 
this is motivated by the 3-dim $Sp(N)$ higher spin theory in
\cite{Anninos:2011ui} but the eventual results appear independent of 
$dS/CFT$ per se. Towards regarding the areas of $dS_4$ complex
extremal surfaces in \cite{Narayan:2015vda,Narayan:2015oka} as
entanglement entropy in the dual Euclidean $CFT_3$ with negative
central charge, it would be interesting to understand if these
features of negative EE in the 2-dim toy investigations here carry
over in the sense here to that case\footnote{Compactifying a free 
massless 3-dim CFT along one direction gives several 2-dim massive 
theories: naively the entanglement entropy summing over modes is 
estimated as (\ref{EEcftd}) using the 2-dim CFT results here, and 
assuming (\ref{EEmassive}) can be established for negative central 
charge as well.}. 
It is worth noting however that while the CFTs discussed here 
do not exhibit any obstruction to a Lorentzian continuation, the
dual CFT in the de Sitter context is intrinsically Euclidean: the
boundary Euclidean time direction implicit in the entanglement
calculations is simply one of the boundary spatial directions along
which there is translation invariance (see \eg\ \cite{Ng:2012xp} for
discussions on a state-operator correspondence in the $dS/CFT$
context).  Relatedly in $dS_3$, the dual is expected to be a 2-dim CFT
with imaginary central charge, with likely new features altogether
(see \eg\ \cite{Balasubramanian:2001nb,Balasubramanian:2002zh} for
aspects of novel hermiticity relations): the CFTs here with negative
central charge are not to be considered as dual to $dS_3$.

It is worth emphasising that the analysis here is formal, suggesting
that a formal generalization of the usual notions of entanglement
entropy for unitary theories can be defined for certain ghost CFTs
with negative central charge. The resulting quantity is negative, as
are the areas of the $dS_4$ complex extremal surfaces reviewed in
sec.~2. Such a negative entanglement entropy does not satisfy
properties of ordinary EE such as strong subadditivity: it has various
odd features, and the physical interpretation is far from clear.
Indeed we call this entanglement entropy only in the sense of an
extension of the usual techniques to this case. Let us recall the area
(\ref{EEdS4strip}) of the codim-2 complex extremal surfaces in $dS_4$
for a strip (and sphere) subregion on a constant boundary Euclidean
time slice studied in \cite{Narayan:2015vda,Narayan:2015oka}. We have\
$S_A \sim -{R_{dS}^2\over G_4} ({V_1\over\epsilon} - {V_1\over l})$\
for a strip of finite width $l$. Then, as discussed in 
\cite{Narayan:2015vda}, the analog of mutual information for two 
disjoint strip subregions $A, B$ defined as\ 
$I[A,B]=S[A]+S[B]-S[A\cup B]$\ is negative definite for $A, B$ 
sufficiently nearby (and vanishes beyond a critical 
separation)\footnote{Entanglement entropy for a single interval 
involves the 2-point function of twist operators in the CFT. 
Multiple intervals may be more interesting: for instance, mutual 
information for two disjoint intervals $A$ and $B$ involves the 
4-point function of twist operators with more intricate dependence 
on the full CFT structure. Relatedly it would be interesting to 
study excited states.}. (The negative sign suggests that $-S[A]$ 
satisfies strong subadditivity.)  Secondly, 
consider two strip subregions of width $l_2$ and $l_1>l_2$\ (for 
intervals with finite widths $l_1, l_2\ll V_1$)\footnote{In the 
CFT context here, we have implicitly assumed that the single interval 
$A$ in question here is finite (although it can be large). The 
presence of the background charge makes it difficult to address 
the question of $S_A$ matching the entanglement entropy of the 
complement of the subsystem in this context.}. Then we see that 
$S(l_1) - S(l_2) = -{R_{dS}^2\over G_4} ({V_1\over l_2} 
- {V_1\over l_1}) < 0$, \ie\ $S(l_1)<S(l_2)$.\ 
Formally the $l\ra\infty$ limit\footnote{It is interesting to ask if 
there are analogs in the bulk de Sitter space context of the background 
charge appearing in our CFT discussions here.} 
gives $S(l)\ra -\infty$ so that any finite $l$ gives $S(l)>S(\infty)$: 
relatedly as $l\ra\epsilon$, we have $S(l)\ra 0^-$. This 
means that a bigger subregion is more ordered than a smaller one, in
contrast with a conventional unitary CFT where $S(l_1)>S(l_2)$ \ie\ a
bigger subregion is more disordered than a smaller one.  Thirdly, the
entropic c-function \cite{Casini:2004bw,Casini:2006es} defined as\
$c(l) = l {dS_A\over dl}$ for unitary 2-dim theories with positive
central charge shows $c(l)$ to be monotically decreasing as the size
$l$ increases, with $c'(l)\leq 0$.  In higher dimensions, this can be
defined as\ $c(l)={l^{d-1}\over V_{d-2}} {dS_A\over dl}$ for
strip-shaped subregions, with $V_{d-2}$ the interface area (see \eg\
\cite{HEEreview}). The finite cutoff-independent part of entanglement
entropy in this regard encodes useful information about the flow
towards long wavelengths. Applying this to the $dS_4$ complex surfaces
area, we have\ $c(l) \equiv {l^2\over V_1} {dS_A\over dl} =
-{R_{dS}^2\over G_4} < 0$ which is negative, reflecting the comments
above, \ie\ as $l$ increases, $S(l)$ decreases.  For more general
asymptotically $dS_4$ spaces, the areas $S_A$ of corresponding 
complex extremal surfaces would again appear to be negative (as noted
in \cite{Narayan:2015vda} for $dS_4$ black branes \cite{Das:2013mfa}).
This would imply that $c'(l) > 0$, \ie\ as the size $l$ increases,
$c(l)$ will increase.  This suggests that new degrees of freedom are
\emph{integrated in}.  This is a little reminiscent of the picture in
\cite{Strominger:2001gp} of time evolution encoded as inverse RG
flow. In the present context, recall that these complex extremal
surfaces were obtained along the path $\tau=iT$, the turning point
being $\tau_*=il$: however each such complex $\tau$ can be mapped to a
corresponding real-valued $\tau$ location in the bulk $dS_4$ space (as
discussed in sec.~2), with $|\tau_*|=l$. Thus increasing size $l$
corresponds to going to larger $|\tau_*|$, \ie\ earlier times in the
past.  It would be interesting to explore this.

Finally, the $dS/CFT$ dictionary $Z_{CFT}=\Psi_{dS}$ suggests that,
unlike in $AdS/CFT$, entanglement entropy in the dual CFT here is 
not the entanglement entropy of bulk fields in $dS$ (see \eg\
\cite{Maldacena:2012xp}). The CFT entanglement entropy, in a replica
formulation, involves ${Z_n\over Z_1^n}$ which naively maps to
${\Psi_n\over\Psi_1^n}$~: relatedly, the bulk density matrix involves 
$\Psi^*\Psi\sim Z^*Z$ which is expected to lead to very different 
entanglement structures. Thus while the areas of the de Sitter
complex extremal surfaces in \cite{Narayan:2015vda,Narayan:2015oka}
might be akin to entanglement entropy in the dual CFT, the bulk
interpretation of these would be interesting to understand better,
perhaps along the lines of
\cite{Lewkowycz:2013nqa,Hartman:2013mia,Faulkner:2013yia}.

\vspace{10mm}
%\newpage

{\footnotesize \noindent {\bf Acknowledgements:}\ \
It is a pleasure to thank 
S. Das, T. Hartman, D. Jatkar, S. Minwalla and A. Sen for 
discussions and correspondence on this work.  I have also benefitted 
from conversations with S. Das, D. Harlow, T. Hartman, R. Loganayagam,
J. Maldacena, G. Mandal, S. Minwalla, S. Mukhi, R. Myers, 
A. Strominger, T. Takayanagi and S. Trivedi on \cite{Narayan:2015vda,
Narayan:2015oka}. I thank the Organizers of the Strings 2015
conference, ICTS and IISc, Bangalore, and the 4th Indian-Israeli
workshop on Quantum Field Theory and String Theory, Goa, as well as
the string theory groups in TIFR, Mumbai, Saha Institute, Kolkata, 
and HRI, Allahabad, for hospitality while this work was in progress. 
This work is partially supported by a grant to CMI from the Infosys 
Foundation.
}

%\vspace{1mm}

\appendix

\section{2-dim CFT and the replica formulation}

Here we review some aspects of the replica formulation of entanglement 
entropy \cite{Calabrese:2004eu,Calabrese:2009qy}, for a 2-dim Euclidean 
CFT with central charge $c$, keeping in mind our context in sec.~3.2. 
Consider a subsystem $A$ comprising a single interval between $x=u$ 
and $x=v>u$ on a time slice $t_E=const$. 
The path integral for $n$ copies of the subsystem is\
%\be
$tr \rho_A^n = \prod_{k=1}^n \langle \Phi_k^u \Phi_k^v \rangle$\
%\ee
where the boundary conditions joining each copy $k$ to its adjacent 
next copy in the replica space are implemented by twist operators 
$\Phi_k^u$ at $w=u$ and $\Phi_k^v$ at $w=v$. The conformal dimensions 
of these twist operators $\Phi_k^u, \Phi_k^v$ can be obtained as in 
\cite{Calabrese:2004eu,Calabrese:2009qy} by using the conformal map 
\be\label{zwConftrans}
z = \Big( {w-u\over w-v} \Big)^{1/n}
\ee
from the replica $w$-space which contains a conical singularity around 
each twist field location, to the replica $z$-space $\BC$. We note that 
the conformal map $w\ra {w-u\over w-v}$ maps the interval points 
$w=u$ and $w=v$ to $0,\ \infty$ respectively: the $z$ transformation 
then maps the $n$ $w$-sheets to a single complex plane $\BC$. \
Near $w\sim u$ and $w\sim v$, we have conical singularities\ 
%\be
$z \sim (w-u)^{1/n}$\ and\ $z \sim (w-v)^{-1/n}$\
%\ee
so that the neighbourhood of these points resembles $z=w^{\pm 1/n}$. 
The metric near these locations is
$dzd{\bar z} = |\del_wz|^2 dw d{\bar w} = {|w|^{-2+2/n}\over n^2} dwd{\bar w}$.
Under a conformal transformation $w\ra z$, the energy-momentum tensor 
transforms as (\ref{TzConfTransf}). The condition (\ref{Tz=0}), \ie\ 
$\langle T(z)\rangle_{\BC} = 0$,\ is equivalent to taking the $z$-plane 
to represent the CFT ground state, 
and so by the state-operator correspondence, equivalent to the insertion 
of the identity operator which corresponds to the ground state.
Now taking expectation values (with ${\cal R}_n$ the $n$-sheeted 
$w$-space), we obtain (\ref{T(w)vev}). For $u, v$ well-separated, \eg\ 
$v\gg u$, the expression (\ref{T(w)vev}) simplifies to\ 
$\langle T(w)\rangle_{{\cal R}_n} \sim\ 
{c (1-{1\over n^2})\over 24} \ {1\over (w-u)^2}$\ which corresponds 
to a conformal transformation $z=(w-u)^{1/n}$ showing a conical 
singularity at $w\sim u$.
The boundary conditions at $u, v$ are equivalent to the insertion of 
twist operators $\Phi_n(u), \Phi_{-n}(v)$ at $w=u, v$ respectively. 
The last expression in (\ref{T(w)vev}) has been written after 
comparing with the standard expression\ 
$\langle T(z) \phi_1(w_1) \phi_2(w_2) \rangle\ =\ \sum_i (
{h_i\over (z-w_i)^2} + {1\over (z-w_i)} {\del\over\del w_i} ) 
\langle \phi_1(w_1) \phi_2(w_2) \rangle$\ for the 3-point function of 
$T(w)$ with two primary operators, \ie\ 
\bea
&& \langle T(w) \Phi_n(u) \Phi_{-n}(v) \rangle\ =\ 
{\Delta_n\over (w-u)^2 (w-v)^2 (v-u)^{2\Delta_n-2} 
({\bar v}-{\bar u})^{2{\bar\Delta}_n}} \nonumber\\
&& \quad = \Big( {\Delta_n\over (w-u)^2} + {\Delta_n\over (w-u)^2} 
+ {1\over (w-u)} {\del\over\del u} + {1\over (w-v)} {\del\over\del v} \Big) 
\langle \Phi_n(u) \Phi_{-n}(v) \rangle\ .
\eea
(We have effectively assumed that $w$ here represents a single sheet 
of the $n$-sheeted $w$-space.) Comparing the second line with 
(\ref{T(w)vev}) leads to the correlation function of the twist operators,
\be\label{twistops}
\langle \Phi_n(u) \Phi_{-n}(v) \rangle\ =\ |v-u|^{-2\Delta_n-2{\bar\Delta}_n}\ ,
\qquad \Delta_n = {c (1-{1\over n^2})\over 24} = {\bar\Delta}_n\ .
\ee
Writing (\ref{T(w)vev}) as\ 
$\langle T(w)\rangle = {\int D\varphi T(w) e^{-S}\over \int D\varphi e^{-S}}$ 
we can write the partition function of the CFT on the $n$-sheeted 
replica space with these boundary conditions as
\be
tr \rho_A^n\ =\ \prod_{k=1}^n \langle \Phi_k(u) \Phi_{-k}(v) \rangle\ 
=\ |v-u|^{c (n-{1\over n})/6}\ ,
\ee
so that the entanglement entropy for the interval is
\be
S^{EE}_A\ 
%=\ \displaystyle{\lim_{n\to 1}}\ \del_n {tr \rho_A^n - 1\over 1-n}\ 
=\ -\displaystyle{\lim_{n\to 1}}\ \del_n tr\rho_A^n\ 
=\ {c\over 3} \log {l\over \epsilon}\ ,
\ee
where $l\equiv |v-u|$ is the length of the interval and $\epsilon$ is 
an ultraviolet cutoff.

The replica formulation above for entanglement entropy appears to
require no input besides the central charge of the CFT: however there
are various implicit assumptions where unitarity has entered. In
sec.~3.2, we have discussed these and corresponding modifications in
the context of the $bc$-ghost system.

\vspace{1mm}

\noindent \emph{Higher dimensional free fields compactified to 2-dim:}\ \
Considering first a massive 2d theory, we imagine the mass $m$ is the only 
scale in the system, which plays the role of the correlation length
$\xi\sim {1\over m}$~. Revisiting the discussion
\cite{Calabrese:2004eu} for entanglement entropy for a single interval
(for simplicity semi-infinite) in this massive theory gives
\be\label{EEmassive}
S_A   %= -\displaystyle{\lim_{n\to 1}}\ \del_n tr \rho_A^n 
%= -\displaystyle{\lim_{n\to 1}}\ \del_n tr \log\rho_A^n 
%= -\displaystyle{\lim_{n\to 1}}\ \del_n \log {Z_n\over Z_1^n} = 
= -{c\over 6} \log (m\epsilon)\ .
\ee
Now following \cite{Ryu:2006ef}, we consider a $d$-dimensional
Euclidean CFT comprising massless free fields. Compactifying this on 
a $(d-2)$-dim torus $T^{d-2}$ gives an effective 2-dim theory on 
$\BR^2$ with many massive fields of mass
\be
m^2=\sum_{i=3}^d k_i^2=\Big({2\pi\over L}\Big)^2\sum_{i=3}^d n_i^2\ ,
\qquad\qquad   k_i={2\pi n_i\over L}\ .
\ee
The correlation length is $\xi\sim {1\over m}$ for modes of mass $m$.
We are considering a strip subsystem with width $l$ along the 
noncompact direction, and stretched along the other directions which 
we compactify. Then the entanglement entropy for this subsystem can be 
estimated by summing over all these modes using the results for 
$l\ll\xi$ and $l\gg\xi$ becomes
\bea
S_A &=& {\displaystyle\sum_{k_3,k_4,\ldots,k_d}^{\xi< l}} {c\over 3} 
\log {\xi\over\epsilon}\ +\
{\displaystyle\sum_{k_3,k_4,\ldots,k_d}^{\xi\geq l}} {c\over 3} 
\log {l\over\epsilon} \nonumber\\
&=& {c\over 3} \left({L\over 2\pi}\right)^{d-2} \left[ \int_{1/l}^{1/\epsilon} 
d^{d-2}k\ (-\log (|k|\epsilon))\ +\ \int_0^{1/l} d^{d-2}k\ \log {l\over\epsilon}
\right]\ .
\eea
Noting $\int k^{d-3}dk \log k = {k^{d-2}\over d-2} (\log k - {1\over d-2})$, 
the above expression shows various cancellations and simplifies as
\be\label{EEcftd}
%S_A &=& {c\over 3} {L^{d-2} \Omega_{d-2}\over (2\pi)^{d-2}} \left[
%{k^{d-2}\over d-2} \Big(\log k - {1\over d-2}\Big)\Big|_{1/\epsilon}^{1/l}\ 
%-\ {k^{d-2}\over d-2} \log\epsilon \Big|_{1/l}^{1/\epsilon}\ +\ 
%{k^{d-2}\over d-2} \log {l\over\epsilon} \Big|_0^{1/l} \right]\ \nonumber\\
S_A = {c\over 3} {\Omega_{d-2}\over (2\pi)^{d-2} (d-2)} \left( 
{L^{d-2}\over\epsilon^{d-2}}\ -\ {L^{d-2}\over l^{d-2}} \right)\ .
\ee
The expression (\ref{EEcftd}) for entanglement entropy obtained by
compactifying a higher dimensional CFT to 2-dim (using
(\ref{EEmassive})) naively appears to be valid beyond the realm of
conventional unitary CFTs.  In particular for negative central charge,
(\ref{EEcftd}) has the same form as the area (\ref{EEdS4strip}) of the
$dS_4$ complex extremal surfaces we reviewed earlier. In this context,
we imagine taking a free 3-dim Euclidean CFT of negative central
charge ${\cal C}\sim -{R_{dS}^2\over G_4}$ (obtained from the
holographic $\langle TT\rangle$ correlators) to be compactified along
one direction giving several 2-dim massive theories. This is simply a
heuristic argument to obtain a rough estimate for the form of
entanglement entropy in a higher dimensional CFT.


\begin{thebibliography}{} %{99}

\footnotesize{

%\cite{Strominger:2001pn}
\bibitem{Strominger:2001pn} 
  A.~Strominger,
  ``The dS / CFT correspondence,''
  JHEP {\bf 0110}, 034 (2001)
  [hep-th/0106113].

%\cite{Witten:2001kn}
\bibitem{Witten:2001kn} 
  E.~Witten,
  ``Quantum gravity in de Sitter space,''
  [hep-th/0106109]).
  %%CITATION = HEP-TH/0106109;%%

%\cite{Maldacena:2002vr}
\bibitem{Maldacena:2002vr}
  J.~M.~Maldacena,
  ``Non-Gaussian features of primordial fluctuations in single field inflationary models,''
  JHEP {\bf 0305}, 013 (2003),\ 
  [astro-ph/0210603].

%\cite{Maldacena:1997re}
\bibitem{Maldacena:1997re}
  J.~M.~Maldacena,
  ``The large N limit of superconformal field theories and supergravity,''
  Adv.\ Theor.\ Math.\ Phys.\  {\bf 2}, 231 (1998)
  [Int.\ J.\ Theor.\ Phys.\  {\bf 38}, 1113 (1999)]
  [arXiv:hep-th/9711200].
  %%CITATION = IJTPB,38,1113;%%

%\cite{Gubser:1998bc}
\bibitem{Gubser:1998bc}
  S.~S.~Gubser, I.~R.~Klebanov and A.~M.~Polyakov,
  ``Gauge theory correlators from non-critical string theory,''
  Phys.\ Lett.\  B {\bf 428}, 105 (1998)
  [arXiv:hep-th/9802109].
  %%CITATION = PHLTA,B428,105;%%

%\cite{Witten:1998qj}
\bibitem{Witten:1998qj}
  E.~Witten,
  ``Anti-de Sitter space and holography,''
  Adv.\ Theor.\ Math.\ Phys.\  {\bf 2}, 253 (1998)
  [arXiv:hep-th/9802150].
  %%CITATION = 00203,2,253;%%

%\cite{Aharony:1999ti}
\bibitem{Aharony:1999ti}
  O.~Aharony, S.~S.~Gubser, J.~M.~Maldacena, H.~Ooguri and Y.~Oz,
  ``Large N field theories, string theory and gravity,''
  Phys.\ Rept.\  {\bf 323}, 183 (2000)
  [arXiv:hep-th/9905111].
  %%CITATION = PRPLC,323,183;%%

%\cite{Strominger:2001gp}
\bibitem{Strominger:2001gp} 
  A.~Strominger,
  ``Inflation and the dS / CFT correspondence,''
  JHEP {\bf 0111}, 049 (2001)
  doi:10.1088/1126-6708/2001/11/049
  [hep-th/0110087].
  %%CITATION = doi:10.1088/1126-6708/2001/11/049;%%

%\cite{Balasubramanian:2001nb}
\bibitem{Balasubramanian:2001nb} 
  V.~Balasubramanian, J.~de Boer and D.~Minic,
  ``Mass, entropy and holography in asymptotically de Sitter spaces,''
  Phys.\ Rev.\ D {\bf 65}, 123508 (2002)
  [hep-th/0110108].
  %%CITATION = HEP-TH/0110108;%%

%\cite{Balasubramanian:2002zh}
\bibitem{Balasubramanian:2002zh} 
  V.~Balasubramanian, J.~de Boer and D.~Minic,
  ``Notes on de Sitter space and holography,''
  Class.\ Quant.\ Grav.\  {\bf 19}, 5655 (2002)
  [Annals Phys.\  {\bf 303}, 59 (2003)]
  [hep-th/0207245].
  %%CITATION = HEP-TH/0207245;%%

%\cite{Harlow:2011ke}
\bibitem{Harlow:2011ke} 
  D.~Harlow and D.~Stanford,
  ``Operator Dictionaries and Wave Functions in AdS/CFT and dS/CFT,''
  arXiv:1104.2621 [hep-th].

%\cite{Anninos:2011ui}
\bibitem{Anninos:2011ui} 
  D.~Anninos, T.~Hartman and A.~Strominger,
  ``Higher Spin Realization of the dS/CFT Correspondence,''
  arXiv:1108.5735 [hep-th].

%\cite{Ng:2012xp}
\bibitem{Ng:2012xp} 
  G.~S.~Ng and A.~Strominger,
  ``State/Operator Correspondence in Higher-Spin dS/CFT,''
  Class.\ Quant.\ Grav.\  {\bf 30}, 104002 (2013)
  [arXiv:1204.1057 [hep-th]].
  %%CITATION = ARXIV:1204.1057;%%

%\cite{Das:2012dt}
\bibitem{Das:2012dt} 
  D.~Das, S.~R.~Das, A.~Jevicki and Q.~Ye,
  ``Bi-local Construction of Sp(2N)/dS Higher Spin Correspondence,''
  JHEP {\bf 1301}, 107 (2013)
  [arXiv:1205.5776 [hep-th]].
  %%CITATION = ARXIV:1205.5776;%%

%\cite{Anninos:2012ft}
\bibitem{Anninos:2012ft} 
  D.~Anninos, F.~Denef and D.~Harlow,
  ``The Wave Function of Vasiliev's Universe - A Few Slices Thereof,''
  Phys.\ Rev.\ D {\bf 88}, 084049 (2013)
  [arXiv:1207.5517 [hep-th]].
  %%CITATION = ARXIV:1207.5517;%%

%\cite{Das:2013qea}
\bibitem{Das:2013qea} 
  D.~Das, S.~R.~Das and G.~Mandal,
  ``Double Trace Flows and Holographic RG in dS/CFT correspondence,''
  arXiv:1306.0336 [hep-th].
  %%CITATION = ARXIV:1306.0336;%%

%\cite{Banerjee:2013mca}
\bibitem{Banerjee:2013mca} 
  S.~Banerjee, A.~Belin, S.~Hellerman, A.~Lepage-Jutier, A.~Maloney, Đj.~đj.~Radičević and S.~Shenker,
  ``Topology of Future Infinity in dS/CFT,''
  JHEP {\bf 1311}, 026 (2013)
  [arXiv:1306.6629 [hep-th]].
  %%CITATION = ARXIV:1306.6629;%%

%\cite{Das:2013mfa}
\bibitem{Das:2013mfa} 
  D.~Das, S.~R.~Das and K.~Narayan,
  ``dS/CFT at uniform energy density and a de Sitter 'bluewall',''
  JHEP {\bf 1404}, 116 (2014)
  [arXiv:1312.1625 [hep-th]].
  %%CITATION = ARXIV:1312.1625;%%

%\cite{Narayan:2015vda}
\bibitem{Narayan:2015vda} 
  K.~Narayan,
  ``de Sitter extremal surfaces,''
  Phys.\ Rev.\ D {\bf 91}, no. 12, 126011 (2015)
  doi:10.1103/PhysRevD.91.126011
  [arXiv:1501.03019 [hep-th]].
  %%CITATION = doi:10.1103/PhysRevD.91.126011;%%

%\cite{Narayan:2015oka}
\bibitem{Narayan:2015oka} 
  K.~Narayan,
  ``de Sitter space and extremal surfaces for spheres,''
  Phys.\ Lett.\ B {\bf 753}, 308 (2016)
  doi:10.1016/j.physletb.2015.12.019
  [arXiv:1504.07430 [hep-th]].
  %%CITATION = doi:10.1016/j.physletb.2015.12.019;%%

%\cite{Ryu:2006bv}
\bibitem{Ryu:2006bv} 
  S.~Ryu and T.~Takayanagi,
  ``Holographic derivation of entanglement entropy from AdS/CFT,''
  Phys.\ Rev.\ Lett.\  {\bf 96}, 181602 (2006)
  [hep-th/0603001].
  %%CITATION = HEP-TH/0603001;%%

%\cite{Ryu:2006ef}
\bibitem{Ryu:2006ef} 
  S.~Ryu and T.~Takayanagi,
  ``Aspects of Holographic Entanglement Entropy,''
  JHEP {\bf 0608}, 045 (2006)
  [hep-th/0605073].
  %%CITATION = HEP-TH/0605073;%%

\bibitem{HRT} 
V.~E.~Hubeny, M.~Rangamani and T.~Takayanagi,
``A Covariant holographic entanglement entropy proposal,'' 
JHEP {\bf 0707} (2007) 062  [arXiv:0705.0016 [hep-th]].
%%CITATION = ARXIV:0705.0016;%%
 
%\cite{Nishioka:2009un}
%\bibitem{Nishioka:2009un} 
\bibitem{HEEreview}
  T.~Nishioka, S.~Ryu and T.~Takayanagi,
  ``Holographic Entanglement Entropy: An Overview,''
  J.\ Phys.\ A {\bf 42}, 504008 (2009)
  doi:10.1088/1751-8113/42/50/504008
  [arXiv:0905.0932 [hep-th]].
  %%CITATION = doi:10.1088/1751-8113/42/50/504008;%%

\bibitem{HEEreview2}
T.~Takayanagi,
  ``Entanglement Entropy from a Holographic Viewpoint,''
  Class.\ Quant.\ Grav.\  {\bf 29} (2012) 153001  [arXiv:1204.2450 [gr-qc]].
  %%CITATION = ARXIV:1204.2450;%%

%\cite{Lewkowycz:2013nqa}
\bibitem{Lewkowycz:2013nqa} 
  A.~Lewkowycz and J.~Maldacena,
  ``Generalized gravitational entropy,''
  JHEP {\bf 1308}, 090 (2013)
  [arXiv:1304.4926 [hep-th]].
  %%CITATION = ARXIV:1304.4926;%%

%\cite{Hartman:2013mia}
\bibitem{Hartman:2013mia} 
  T.~Hartman,
  ``Entanglement Entropy at Large Central Charge,''
  arXiv:1303.6955 [hep-th].
  %%CITATION = ARXIV:1303.6955;%%

%\cite{Faulkner:2013yia}
\bibitem{Faulkner:2013yia} 
  T.~Faulkner,
  ``The Entanglement Renyi Entropies of Disjoint Intervals in AdS/CFT,''
  arXiv:1303.7221 [hep-th].
  %%CITATION = ARXIV:1303.7221;%%

%\cite{Sato:2015tta}
\bibitem{Sato:2015tta} 
  Y.~Sato,
  ``Comments on Entanglement Entropy in the dS/CFT Correspondence,''
  Phys.\ Rev.\ D {\bf 91}, no. 8, 086009 (2015)
  [arXiv:1501.04903 [hep-th]].
  %%CITATION = ARXIV:1501.04903;%%

%\cite{Belavin:1984vu}
\bibitem{Belavin:1984vu} 
  A.~A.~Belavin, A.~M.~Polyakov and A.~B.~Zamolodchikov,
  ``Infinite Conformal Symmetry in Two-Dimensional Quantum Field Theory,''
  Nucl.\ Phys.\ B {\bf 241}, 333 (1984).
  doi:10.1016/0550-3213(84)90052-X
  %%CITATION = doi:10.1016/0550-3213(84)90052-X;%%

%\cite{Holzhey:1994we}
\bibitem{Holzhey:1994we} 
  C.~Holzhey, F.~Larsen and F.~Wilczek,
  ``Geometric and renormalized entropy in conformal field theory,''
  Nucl.\ Phys.\ B {\bf 424}, 443 (1994)
  [hep-th/9403108].
  %%CITATION = HEP-TH/9403108;%%

%\cite{Calabrese:2004eu}
\bibitem{Calabrese:2004eu} 
  P.~Calabrese and J.~L.~Cardy,
  ``Entanglement entropy and quantum field theory,''
  J.\ Stat.\ Mech.\  {\bf 0406}, P06002 (2004)
  [hep-th/0405152].
  %%CITATION = HEP-TH/0405152;%%

%\cite{Calabrese:2009qy}
\bibitem{Calabrese:2009qy} 
  P.~Calabrese and J.~Cardy,
  ``Entanglement entropy and conformal field theory,''
  J.\ Phys.\ A {\bf 42}, 504005 (2009)
  doi:10.1088/1751-8113/42/50/504005
  [arXiv:0905.4013 [cond-mat.stat-mech]].
  %%CITATION = doi:10.1088/1751-8113/42/50/504005;%%

\bibitem{polchinskiTextBk}
J.~Polchinski, String Theory, Vol. 1,2. 
Cambridge University Press (1998).

%\cite{Blumenhagen:2013fgp}
\bibitem{Blumenhagen:2013fgp} 
  R.~Blumenhagen, D.~Lüst and S.~Theisen,
  ``Basic concepts of string theory,'' Springer  (2013).
%  doi:10.1007/978-3-642-29497-6.
  %%CITATION = doi:10.1007/978-3-642-29497-6;%%

%\cite{Friedan:1985ge}
\bibitem{Friedan:1985ge} 
  D.~Friedan, E.~J.~Martinec and S.~H.~Shenker,
  ``Conformal Invariance, Supersymmetry and String Theory,''
  Nucl.\ Phys.\ B {\bf 271}, 93 (1986).
  doi:10.1016/S0550-3213(86)80006-2
  %%CITATION = doi:10.1016/S0550-3213(86)80006-2;%%

%\cite{Saleur:1991hk}
\bibitem{Saleur:1991hk} 
  H.~Saleur,
  ``Polymers and percolation in two-dimensions and twisted N=2 supersymmetry,''
  Nucl.\ Phys.\ B {\bf 382}, 486 (1992)
  doi:10.1016/0550-3213(92)90657-W
  [hep-th/9111007].
  %%CITATION = doi:10.1016/0550-3213(92)90657-W;%%

%\cite{Kausch:1995py}
\bibitem{Kausch:1995py} 
  H.~G.~Kausch,
  ``Curiosities at c = -2,''
  hep-th/9510149.
  %%CITATION = HEP-TH/9510149;%%

%\cite{Kausch:2000fu}
\bibitem{Kausch:2000fu} 
  H.~G.~Kausch,
  ``Symplectic fermions,''
  Nucl.\ Phys.\ B {\bf 583}, 513 (2000)
%  doi:10.1016/S0550-3213(00)00295-9
  [hep-th/0003029].
  %%CITATION = doi:10.1016/S0550-3213(00)00295-9;%%

%\cite{Flohr:2001zs}
\bibitem{Flohr:2001zs} 
  M.~Flohr,
  ``Bits and pieces in logarithmic conformal field theory,''
  Int.\ J.\ Mod.\ Phys.\ A {\bf 18}, 4497 (2003)
  doi:10.1142/S0217751X03016859
  [hep-th/0111228].
  %%CITATION = doi:10.1142/S0217751X03016859;%%

%\cite{Krohn:2002gh}
\bibitem{Krohn:2002gh} 
  M.~Krohn and M.~Flohr,
  ``Ghost systems revisited: Modified Virasoro generators and logarithmic conformal field theories,''
  JHEP {\bf 0301}, 020 (2003)
  doi:10.1088/1126-6708/2003/01/020
  [hep-th/0212016].
  %%CITATION = doi:10.1088/1126-6708/2003/01/020;%%

%\cite{Gurarie:1993xq}
\bibitem{Gurarie:1993xq} 
  V.~Gurarie,
  ``Logarithmic operators in conformal field theory,''
  Nucl.\ Phys.\ B {\bf 410}, 535 (1993)
  doi:10.1016/0550-3213(93)90528-W
  [hep-th/9303160].
  %%CITATION = doi:10.1016/0550-3213(93)90528-W;%%

%\cite{Gurarie:1997dw}
\bibitem{Gurarie:1997dw} 
  V.~Gurarie, M.~Flohr and C.~Nayak,
  ``The Haldane-Rezayi quantum Hall state and conformal field theory,''
  Nucl.\ Phys.\ B {\bf 498}, 513 (1997)
  doi:10.1016/S0550-3213(97)00351-9
  [cond-mat/9701212].
  %%CITATION = doi:10.1016/S0550-3213(97)00351-9;%%

%\cite{Gurarie:2013tma}
\bibitem{Gurarie:2013tma} 
  V.~Gurarie,
  ``Logarithmic operators and logarithmic conformal field theories,''
  J.\ Phys.\ A {\bf 46}, 494003 (2013)
  doi:10.1088/1751-8113/46/49/494003
  [arXiv:1303.1113 [cond-mat.stat-mech]].
  %%CITATION = doi:10.1088/1751-8113/46/49/494003;%%

%\cite{LeClair:2006kb}
\bibitem{LeClair:2006kb} 
  A.~LeClair,
  ``Quantum critical spin liquids, the 3D Ising model, and conformal field theory in 2+1 dimensions,''
  cond-mat/0610639.
  %%CITATION = COND-MAT/0610639;%%

%\cite{LeClair:2007iy}
\bibitem{LeClair:2007iy} 
  A.~LeClair and M.~Neubert,
  ``Semi-Lorentz invariance, unitarity, and critical exponents of symplectic fermion models,''
  JHEP {\bf 0710}, 027 (2007)
  doi:10.1088/1126-6708/2007/10/027
  [arXiv:0705.4657 [hep-th]].
  %%CITATION = doi:10.1088/1126-6708/2007/10/027;%%

\bibitem{AreaLaw1}
  L.~Bombelli, R.~K.~Koul, J.~Lee and R.~D.~Sorkin,
  ``A Quantum Source of Entropy for Black Holes,''
  Phys.\ Rev.\ D {\bf 34} (1986) 373.
  %% CITATION = PHRVA,D34,373;%%

\bibitem{AreaLaw2}
   M.~Srednicki,
   ``Entropy and area,''
   Phys.\ Rev.\ Lett.\  {\bf 71} (1993) 666  [hep-th/9303048].
   %% CITATION = HEP-TH/9303048;%%

%\cite{Casini:2005rm}
\bibitem{Casini:2005rm} 
  H.~Casini, C.~D.~Fosco and M.~Huerta,
  ``Entanglement and alpha entropies for a massive Dirac field in two dimensions,''
  J.\ Stat.\ Mech.\  {\bf 0507}, P07007 (2005)
%  doi:10.1088/1742-5468/2005/07/P07007
  [cond-mat/0505563].
  %%CITATION = doi:10.1088/1742-5468/2005/07/P07007;%%

%\cite{Casini:2009sr}
\bibitem{Casini:2009sr} 
  H.~Casini and M.~Huerta,
  ``Entanglement entropy in free quantum field theory,''
  J.\ Phys.\ A {\bf 42}, 504007 (2009)
  doi:10.1088/1751-8113/42/50/504007
  [arXiv:0905.2562 [hep-th]].
  %%CITATION = doi:10.1088/1751-8113/42/50/504007;%%

%\cite{Dixon:1986qv}
\bibitem{Dixon:1986qv} 
  L.~J.~Dixon, D.~Friedan, E.~J.~Martinec and S.~H.~Shenker,
  ``The Conformal Field Theory of Orbifolds,''
  Nucl.\ Phys.\ B {\bf 282}, 13 (1987).
  doi:10.1016/0550-3213(87)90676-6
  %%CITATION = doi:10.1016/0550-3213(87)90676-6;%%

%\cite{Bianchini:2014uta}
\bibitem{Bianchini:2014uta} 
  D.~Bianchini, O.~A.~Castro-Alvaredo, B.~Doyon, E.~Levi and F.~Ravanini,
  ``Entanglement Entropy of Non Unitary Conformal Field Theory,''
  J. Phys. A {\bf 48}, no. 4, 04FT01 (2015)
%  doi:10.1088/1751-8113/48/4/04FT01
  [arXiv:1405.2804 [hep-th]].
  %%CITATION = doi:10.1088/1751-8113/48/4/04FT01;%%

%\cite{Caraglio:2008pk}
\bibitem{Caraglio:2008pk} 
  M.~Caraglio and F.~Gliozzi,
  ``Entanglement Entropy and Twist Fields,''
  JHEP {\bf 0811}, 076 (2008)
  [arXiv:0808.4094 [hep-th]].
  %%CITATION = ARXIV:0808.4094;%%

%\cite{Casini:2004bw}
\bibitem{Casini:2004bw} 
  H.~Casini and M.~Huerta,
  ``A Finite entanglement entropy and the c-theorem,''
  Phys.\ Lett.\ B {\bf 600}, 142 (2004)
  doi:10.1016/j.physletb.2004.08.072
  [hep-th/0405111].
  %%CITATION = doi:10.1016/j.physletb.2004.08.072;%%

%\cite{Casini:2006es}
\bibitem{Casini:2006es} 
  H.~Casini and M.~Huerta,
  ``A c-theorem for the entanglement entropy,''
  J.\ Phys.\ A {\bf 40}, 7031 (2007)
  doi:10.1088/1751-8113/40/25/S57
  [cond-mat/0610375].
  %%CITATION = doi:10.1088/1751-8113/40/25/S57;%%

%\cite{Maldacena:2012xp}
\bibitem{Maldacena:2012xp} 
  J.~Maldacena and G.~L.~Pimentel,
  ``Entanglement entropy in de Sitter space,''
  JHEP {\bf 1302}, 038 (2013)
  [arXiv:1210.7244 [hep-th]].
  %%CITATION = ARXIV:1210.7244;%%


}
\end{thebibliography}
\end{document}